\documentclass[journal]{IEEEtran}
\usepackage{algorithmic}
\usepackage{algorithm}
\usepackage{textcomp}
\usepackage{booktabs}
\usepackage{url}
\usepackage{graphicx}
\usepackage{cite}
\usepackage{amsmath,amssymb,amsfonts}
\usepackage{xcolor}
\begin{document}

\title{Digital Twin–Enabled Mobility-Aware Cooperative Caching in Vehicular Edge Computing}

\author{Jiahao Zeng, Zhenkui Shi$^{*}$, Chunpei Li, Mengkai Yan, Hongliang Zhang, Sihan Chen, Xiantao Hu$^{*}$, Xianxian Li

\thanks{This work was supported by the Key Lab of Education Blockchain and Intelligent Technology, Ministry of Education (EBME25-F-16), Guangxi Key Lab of Brain-inspired Computing and Intelligent Chips (BCIC-25-K1), National Natural Science Foundation of China (U21A20474), and Guangxi Natural Science Foundation (2025GXNSFBA069295). Additional support was provided by the Postdoctoral Fellowship Program of CPSF (GZC20251060), China Postdoctoral Science Foundation (2025MD784113), and the Basic Ability Enhancement Program for Young and Middle-aged Teachers of Guangxi (2025KY0105). \textit{(Corresponding author: Zhenkui Shi, Xiantao Hu.)}}
\thanks{Jiahao Zeng, Zhenkui Shi, Chunpei Li, Xiantao Hu and Xianxian Li 
are with the Key Laboratory of Education Blockchain and Intelligent Technology, Ministry of Education, Guangxi Key Laboratory of Brain-inspired Computing and Intelligent Chips, the Guangxi Key Laboratory of Multi-Source Information Mining and Security, University Engineering Research Center of Educational Intelligent Technology, and School of Computer Science and Engineering, Guangxi Normal University, Guilin 541004, China (e-mail: zengjh@stu.gxnu.edu.cn; shizhenkui@gxnu.edu.cn; licp@gxnu.edu.cn; huxiantao481@gmail.com; lixx@gxnu.edu.cn).}
\thanks{Mengkai Yan, Hongliang Zhang and Sihan Chen  are with PCA Lab, Key Lab of Intelligent Perception and Systems for High-Dimensional Information of Ministry of Education, School of Computer Science and Engineering, Nanjing University of Science and Technology. (e-mail: ymk@njust.edu.cn, zhang1hongliang@njust.edu.cn, melody@njust.edu.cn).}
}
\markboth{Journal of \LaTeX\ Class Files,~Vol.~14, No.~8, August~2021}%
{Shell \MakeLowercase{\textit{et al.}}: A Sample Article Using IEEEtran.cls for IEEE Journals}

\IEEEpubid{0000--0000/00\$00.00~\copyright~2021 IEEE}


\maketitle

\begin{abstract}

With the advancement of vehicle-to-vehicle (V2V) ad hoc networks and wireless communication technologies, mobile edge caching has become a key enabler for enhancing network performance and user experience. However, traditional federated learning–based collaborative caching approaches in vehicular scenarios suffer from inadequate client selection mechanisms and limited prediction accuracy, which result in suboptimal cache hit ratios and increased content transmission latency. To address these challenges, we propose a Digital Twin–based Asynchronous Federated Learning–driven Predictive Edge Caching with Deep Reinforcement Learning (DAPR) framework. DAPR employs an intelligent client selection strategy based on asynchronous federated learning, which leverages mobility prediction and data quality assessment to avoid selecting highly mobile clients or clients with low-quality data, thereby significantly improving model convergence efficiency. In addition, we design a GRU–VAE prediction model that uses a Variational Autoencoder (VAE) to capture latent data distribution features and Gated Recurrent Units (GRUs) to model temporal dependencies, thereby substantially enhancing the accuracy of content request prediction. The predicted content popularities are then fed into a deep reinforcement learning–driven caching decision engine to dynamically optimize edge caching resource allocation. Extensive experiments demonstrate that DAPR achieves superior performance in terms of average reward, cache hit ratio, and transmission latency, thereby effectively improving the overall efficiency of vehicular edge caching systems.

\end{abstract}


\section{Introduction}
\IEEEPARstart{V}ehicle-to-Everything (V2X) networks are foundational technologies for intelligent transportation systems and are currently undergoing a transformative evolution toward 5G/6G network architectures \cite{9454594,9895362}. During this evolution, achieving efficient content distribution and cache management in highly dynamic vehicular environments has become a key research priority \cite{10123387,10423916}. The inherent characteristics of Vehicular Ad Hoc Networks (VANETs)—including high mobility, intermittent connectivity, and heterogeneous resource availability—require sophisticated cooperative caching methods to adapt to rapidly changing network conditions \cite{9770095,8809280}.

Traditional centralized cache management methods have proved inadequate for the distributed and highly dynamic nature of vehicular networks \cite{10420495,10295979}. Federated learning (FL) has therefore attracted significant attention as a promising approach to collaborative vehicular caching, as it enables distributed model training while preserving data privacy \cite{10.1145/3229543.3229555,9344584,FIROUZJAEE2024110142}. By allowing vehicles and roadside units (RSUs) to collaboratively train predictive models without sharing raw data, FL leverages collective knowledge within the network to support intelligent caching decisions \cite{9552606,10745831,QIAN2024111795,10681663}. 

Despite these advantages, existing FL-based collaborative caching schemes face two fundamental challenges that limit their practical effectiveness: 1) ineffective client selection, which leads to degraded training performance. Traditional FL typically employs random selection or strategies based on simple metrics (e.g., computational capability) to determine client participation \cite{10745831,QIAN2024111795}. In vehicular environments, however, the high mobility of vehicles can cause selected clients to leave the communication range or experience severe network degradation during a training cycle, which leads to training interruptions and resource waste. In addition, significant heterogeneity in data quality across vehicles means that indiscriminately selecting clients with poor-quality data can severely impair the convergence and performance of the global model \cite{10681663}. 2) limited content prediction accuracy, which constrains improvements in cache hit ratio. While existing approaches based on long short-term memory (LSTM) networks or simple time prediction models have shown promise, they may face limitations in capturing complex spatio-temporal correlations and user behavior patterns under highly dynamic vehicular environments. These methods are particularly inadequate for handling long-term sequence dependencies and uncertainties in data distribution, which leads to limited prediction accuracy and directly weakens the effectiveness of caching decisions \cite{TAO2022372,10.1145/3711682,huang2025generative}.
\IEEEpubidadjcol

To overcome these limitations, we propose a mobility-aware cooperative caching framework that integrates digital twin technology. The framework is organized as a three-tier architecture: physical, digital twin, and intelligent decision layers. The digital twin layer enables real-time mapping of physical network states and environmental context; the asynchronous federated learning (AFL) layer performs distributed model training under privacy constraints; and the deep reinforcement learning (DRL) layer optimizes dynamic caching resources. The asynchronous federated learning layer incorporates an intelligent client selection mechanism based on mobility prediction and data quality assessment, resolving training stability issues. The digital twin layer collaborates with the asynchronous federated learning layer to support a GRU-VAE hybrid prediction architecture. This architecture employs VAE to capture the latent distribution of data and GRU to model temporal dependencies, enabling precise content request prediction. Prediction results are fed into a deep reinforcement learning-driven caching decision engine, ultimately achieving dynamic optimisation of caching resources. Through hierarchical coordination across modules, this integrated framework systematically resolves the inefficiencies in client selection, insufficient prediction accuracy, and decision-making lag inherent in traditional approaches. 

The main contributions of this paper are summarized as follows.
\begin{itemize}
    \item We propose a new framework that combines digital twin technology with improved asynchronous federated learning to address the limitations of traditional federated learning methods in the context of vehicular collaborative caching, including imperfect client selection mechanisms and insufficient prediction accuracy.
    \item We designed a client selection mechanism based on mobility prediction and data quality assessment, which avoids selecting clients with high mobility or low quality, thereby significantly improving model convergence efficiency in vehicular environments.
    \item We built a predictive model that combines VAE, which captures latent data distribution features, and GRU, which models time dependencies, greatly improving the accuracy of content request predictions and feeding the results back to a deep reinforcement learning-driven caching decision engine to enable dynamic resource allocation.
\end{itemize}

The rest of the paper is organized as follows. In Section \ref{Se:2}, we review the related work. In Section \ref{Se:3}, we briefly introduce the system model. In Section \ref{se:4}, we detail the design of our DAPR scheme. In Section \ref{Se:5}, we test the performance of the proposed scheme, and then in Section \ref{Se:6}, we summarize the work of this paper.

\section{Related work}\label{Se:2}
We review existing content caching schemes, including traditional caching optimization, federated learning methods, and deep reinforcement learning-based content caching schemes.

\subsection{Traditional Cache Optimization and Federated Learning Methods}
Early vehicular caching solutions relied on traditional optimization techniques. Yao et al. \cite{8854137} applied game theory methods, while Neetu et al. \cite{10272617} proposed a mobile terminal-based strategy for the Vehicle Content Center Network (VCCN). However, due to their passive nature and inability to predict mobility patterns, these methods struggle to adapt in dynamic vehicular environments. Additionally, their centralized optimization approaches have proved computationally intensive and difficult to scale as network size increases, highlighting the need for distributed learning methods that can leverage local intelligence while maintaining global coordination.

Recently, FL methods have emerged to address privacy issues in vehicle collaborative caching. Mondal et al. \cite{9773312} introduced the CALM framework for service-as-a-service environments, while Zhou et al. \cite{10122980} proposed Quality of Experience (QoE)-oriented optimization. 
Gu et al. \cite{10745538} innovatively combined federated self-supervised learning (FSSL) with motion-blur resistance mechanisms, providing an effective paradigm for addressing the coordinated challenge of ‘privacy protection – blur resistance – resource optimization’ in V2X networks. Despite these advances, existing federated learning approaches suffer from significant limitations: inadequate client selection mechanisms that neglect vehicle mobility and data quality heterogeneity\cite{10198898,9181482}. When selected vehicles fall out of RSU coverage length or provide low-quality data\cite{10745831, QIAN2024111795}, selection strategies based on randomness or simple metrics often lead to training interruptions. This subsequently reduces model convergence speed and results in computational resource wastage. These limitations hinder practical deployment within vehicle edge computing environments, highlighting the necessity for designing mobile-aware and data-quality-sensitive distributed learning approaches.

\subsection{Deep Reinforcement Learning-based Content Caching Schemes}

Deep reinforcement learning (DRL) has established itself as the mainstream approach in the field of dynamic cache allocation, with numerous research efforts building upon and extending its foundations.

Early research efforts primarily focused on directly applying standard DRL algorithms for cache decision-making. Zhang et al. \cite{9399641} pioneered the application of Deep Deterministic Policy Gradient (DDPG) for RSU caching, treating popularity as a static environmental state; Yang et al. \cite{9139263} employed stochastic Q-learning within the NOMA-MEC framework. However, these approaches rely on short-sighted state observations, neglecting the temporal correlation of content requests, resulting in suboptimal decision-making within the highly dynamic in-vehicle environment.

Subsequent research endeavours sought to enhance DRL performance by incorporating predictive models. Wang et al. \cite{9344584} combined LSTMs with DRL (CCCRP), Qu et al. \cite{10285090} employed GRUs for time-varying MDP modelling, and Wu et al. \cite{10173652}  further proposed the CMCF framework for joint optimisation. Although LSTM architectures can capture short-term dependencies, they struggle to handle the inherent long-term spatio-temporal correlations and data distribution uncertainties within the vehicular environment. While GRUs \cite{10285090} offer improved efficiency and recent studies have even employed Transformers \cite{10843678} to model long-range dependencies, these deterministic models still face a key challenge: they yield point estimates that do not account for the contingent uncertainty inherent in user requests.

Recent studies have begun addressing specific issues \cite{10847910,11023535}. Wu et al. \cite{9944845} proposed a CAFR system employing federated DRL for privacy protection, whilst Zhang et al. \cite{8896914} proposed a socially aware content scheduling scheme. This scheme optimizes local and cross-regional content distribution strategies using deep reinforcement learning algorithms, thereby improving content scheduling efficiency. However, most of these approaches remain reactive in nature, lacking the capacity for proactive simulation of future network conditions. This results in inadequate adaptability to sudden changes in traffic patterns.

\subsection{Digital Twin Technology in Vehicular Edge Caching}
Digital twin (DT) technology has emerged as a transformative paradigm for real-time network optimization by creating high-fidelity virtual replicas of physical entities \cite{TAO2022372,10.1145/3711682}. In vehicular edge computing, digital twins enable dynamic mapping of physical network states into a virtual space for predictive simulation and decision-making. Early applications focused on macro-level traffic modeling and infrastructure management, but recent works have begun exploring its potential for content caching optimization\cite{10521555,10621520, JIHAD2026108144, 10720851,9399641}.

Bandyopadhyay et al. \cite{10681447} have integrated a connected autonomous vehicle system that combines digital twins with non-orthogonal multiple access (NOMA), aiming to address service caching and task offloading optimisation challenges under constrained edge server cache capacity. Li et al. \cite{10959046} have developed a data-driven in-vehicle edge caching and offloading framework to address the complex resource management challenges arising from vehicle mobility. Xu et al. \cite{10720851} proposed the lightweight blockchain consensus mechanism d2BFT (migrated to DT virtual space for execution) and constructs a joint blockchain-communication-computing-caching optimisation problem, solved via a multi-agent double actor-observer algorithm. Chen et al. \cite{10521555} proposed the SACDT-D3 algorithm, which models vehicle states through DT, forms an in-vehicle cloud via semantic similarity clustering, and employs a Dueling Double Deep Q-Network (D3QN) combined with content popularity to implement caching decisions. However, the majority of existing research on digital twins typically employs them for offline analysis rather than as real-time decision support systems continuously feeding contextual information to caching algorithms. Furthermore, the synergistic relationship between digital twins and asynchronous federated learning remains under-explored, particularly concerning how mobile patterns in twin predictions can intelligently select stable clients while balancing their contributions.

These identified gaps have prompted us to propose the DAPR framework, which addresses these limitations through the following approaches: 1) A mobile-aware asynchronous federated learning method with intelligent client selection based on trajectory prediction and data quality assessment. 2) A novel GRU-VAE hybrid prediction architecture that combines the generative capability of variational autoencoders (VAE) to capture latent data distribution features with the temporal modeling advantages of gated recurrent units (GRU). 3) A comprehensive digital twin-assisted real-time decision-making mechanism that provides accurate environmental perception and contextual information to support optimized cache deployment decisions.

\begin{figure*}[t]
\centering 
\includegraphics[width=1\textwidth]{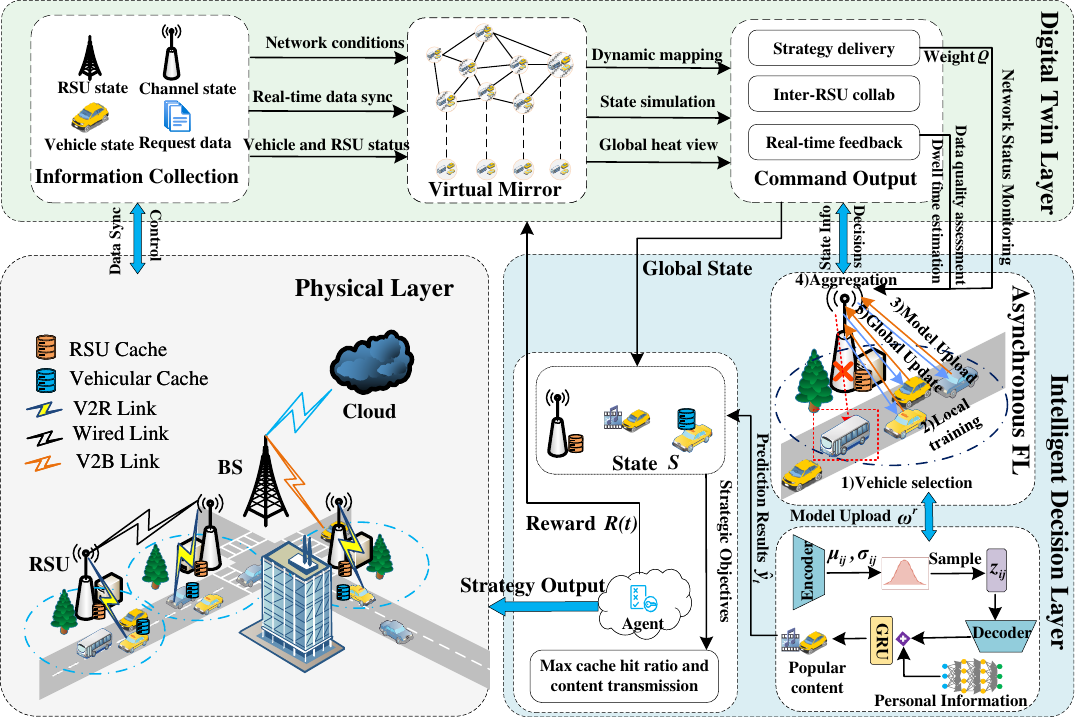} 
\caption{ Schematic of digital twin-driven edge caching for vehicle networking.} 
\label{fig_1} 
\end{figure*}

\section{System Model}\label{Se:3}
\subsection{Preliminary}
As shown in Fig.~\ref{fig_1}, the digital twin-driven on-board edge caching system is divided into the following three main layers:

\textbf{Physical Layer}. The physical layer encompasses the tangible components of the vehicular network, including the Base Station (BS), RSUs, vehicles, and cloud computing resources. The BS provides extensive coverage across the service area. RSUs are positioned at strategic locations to offer edge computing services and store frequently requested content. Vehicles are equipped with onboard units that cache personalized content and collect data such as user requests, location information, and network status. This data is anonymized and transmitted to RSUs or the BS through various communication links, including Vehicle-to-RSU (V2R), wired, and Vehicle-to-Base Station (V2B) links.

\textbf{Digital Twin Layer}. The digital twin layer achieves real-time mapping and simulation optimisation by creating virtual representations of physical layer entities. This layer comprises an information gathering module, virtual representations, global heatmap generation, and command output modules. The information gathering module synchronises physical layer data; virtual representations construct dynamic models of network conditions; the global heatmap provides an overview of content requests; while the command output module disseminates optimisation strategies to the physical layer and intelligent decision-making layer.

\textbf{Intelligent Decision Layer}. The intelligent decision layer optimises caching strategies through AFL, DRL and content popularity prediction. AFL enables distributed model training whilst safeguarding data privacy. DRL agents learn optimal caching strategies based on models aggregated via AFL, whilst content popularity prediction forecasts future demand using historical data.

Specifically, vehicles and RSUs within the physical layer collect and synchronise user request and network status data to the digital twin layer. This layer constructs virtual representations of physical entities, enabling dynamic mapping and simulation optimisation to provide the intelligent decision-making layer with precise environmental awareness and contextual information. Subsequently, the intelligent decision layer utilises an AFL mechanism to aggregate model updates from distributed vehicles while safeguarding user privacy. Through its DRL module, it designs caching strategies based on multi-dimensional rewards. These strategies are then translated by the digital twin layer into execution instructions for the physical layer, optimising cache hit rates and reducing content transmission latency. The entire framework forms a closed-loop system capable of adapting to dynamic changes within the vehicle network, achieving dynamic optimisation of edge caching resource allocation.
The whole system forms a three-layer closely collaborating architecture to build an efficient and secure vehicular edge caching system. Key symbols and their definitions are listed in Table ~\ref{tab_1}.

\begin{table}[htbp]
    \caption{Symbols and Definitions}
    \begin{center}
        \begin{tabular}{@{}p{1.5cm}p{6cm}@{}}
        \toprule
        \textbf{Symbols} & \textbf{Definition} \\ \midrule
        $v_i$ & Travel speed of vehicle $i$  \\ 
        $\rho_i$ & RSU $i$ range vehicle density \\ 
        $L_s$ & RSU coverage length  \\ 
        $q_i$ & Cached content $i$ \\ 
        $W_{i,j}$ & Bandwidth between vehicle $i$ and RSU $j$\\ 
        $d_i$ & Cache Content Size $i$ \\ 
        $P_i$ & The transmission power of the vehicle $i$ \\ 
        $\beta$ & Path loss factor \\ 
        $\sigma^2_{i,j}$ & Gaussian white noise \\ 
        $c_t$ & Contextual characterization of  $t$ time slots \\ 
        $I_i$ &  The vehicle position information at time $t$ \\ 
        $T(r)$ & Duration of the round $r$\\ 
        $\omega^r$ & Global model for round $r$\\ 
        $\alpha_1$ & Data weights \\ 
        $\alpha_2$ & Vehicle location weights \\ 
        $n_i$ & Content $i$ Number of visits \\
        \bottomrule
        \label{tab_1}
        \end{tabular}
    \end{center}
\end{table}

\subsection{Digital Twin Model}
The Digital Twin Layer is implemented with three interconnected modules, each featuring detailed digital twin mechanisms for precise physical-virtual synchronization and simulation-driven optimization. The Information Collection Module aggregates real-time data from the Physical Layer, including RSU/infrastructure state, communication channel conditions, vehicle mobility (position, speed, dwell time), and user request patterns. This data is transmitted securely (e.g., via 5G URLLC) and validated to ensure privacy and integrity.

The Virtual Mirror Module leverages the collected data to construct a high-fidelity virtual replica of the vehicular edge network. It boasts dynamic topology mapping, which updates the virtual network topology in real time, such as the addition or removal of vehicles and changes in RSU coverage. It also 
incorporates a hybrid state-prediction module to forecast future states, including vehicle mobility and network congestion risks. Additionally, it generates a global heat view by aggregating content request data across RSUs to build a spatio-temporal heatmap, quantifying content popularity with metrics like request frequency, geographic distribution, and temporal decay rate, updated every 100 ms to capture real-time demand shifts.

The command output module supports bidirectional interaction for policy execution and status feedback. It receives optimised caching policies from the intelligent decision layer and distributes these policies to target RSUs/vehicles via priority-driven communication, ensuring critical policies are delivered in a timely manner. This module facilitates collaborative caching by sharing a global heatmap view and virtual mirroring between roadside units. It also provides real-time status feedback to the AFL, monitoring network health and data quality while transmitting key metrics to the intelligent decision layer. This feedback enables dynamic adjustment of client selection and model training parameters for federated learning, ensuring its stability in highly mobile scenarios.

\subsection{Vehicle Movement Model}
To simulate real traffic scenarios, this paper proposes a dynamic vehicle movement model based on traffic flow theory \cite{THONHOFER201832,khan2019macroscopic}, in order to characterize the velocity distribution of a vehicle in relation to its density in the area covered by a roadside unit. 
This model is integrated into the digital twin layer to predict vehicle movement behaviour in real time, supporting two core functions: 1) client selection stability in asynchronous federated learning (Section \ref{se:4b}); 2) context-aware popularity prediction (Section \ref{se:4a}). Unlike simple random walk models, this approach captures the collective behaviour of vehicle groups under varying traffic densities, providing a robust and reliable foundation for edge caching decisions. 

At the commencement of each communication round $r$, the RSU calculates the local vehicle density based on the instantaneous position distribution of N vehicles within the coverage area $\rho_i^r$:

\begin{equation}
    \rho_i^r=\frac{N}{L_s},
\end{equation}
where $L_s$ is the RSU coverage length and $N$ is the number of vehicles in the coverage area. This density metric reflects the current level of congestion in the area: $\rho_i^r\to0 $indicates free-flow conditions, while $\rho_i^r\to\rho_{max}$ indicates severe congestion.

The average velocity $v_i^r$ of vehicle $V_i$ in round $r$ is derived according to the Greenshields linear model.

\begin{equation}
    v_i^r = v_f(1-\frac{\rho_i^r}{\rho_{max}}),
\end{equation}
where $v_f$ denotes free-flow velocity (maximum speed without congestion), $\rho_{max}$  represents congestion density, and $(1-\rho_i^r/\rho_{max})$ indicates the speed reduction factor attributable to congestion.

The most critical aspect of this framework is predicting the vehicle's remaining dwell time $T_{stay}$ within the RSU coverage area. $T_{stay}$ can be calculated from the vehicle's current distance travelled $L_i^r$ and predicted speed $v_i^r$\cite{9944845,liu2022learning}:

\begin{equation}
    T_{stay}=\frac{L_s-L_i^r}{v_i^r}.
\end{equation}

Vehicle $V_i$ shall only be selected as a federated learning client when the following stability conditions are satisfied:

\begin{equation}
    T_{stay}>T_{train}+T_{trans},
\end{equation}
where $T_{train}$ denotes the local training duration (positively correlated with data volume and model complexity), and $T_{trans}$ represents the model upload latency (dependent on the uplink rate). This condition ensures selected vehicles can fully participate in training rounds, preventing lagging behind or interruptions.

Then, based on the predicted speed and lap duration $T(r)$, the vehicle's position for the next lap can be updated:
\begin{equation}
    L_i^{r+1}=L_i^r+v_i^r\cdot T(r).
\end{equation}

When the digital twin predicts that $L_i^{r+1}>L_s$, it triggers the cross-RSU handover mechanism, smoothly migrating the training task to the next RSU.

\subsection{Communication Model}
Due to the limited resources of each RSU, it is not possible to cache all the popular contents, and if we only rely on the local cache, when the vehicle requests contents that are not in the local RSU, it will result in cache misses and the need to fetch the contents from a farther place. Therefore, we use $q_i=\{i,n_i,d_i\} $ for the cached content, where $i$ is the content id, $n_i$ denotes the number of times the content has been accessed, and $d_i$ is the size of the content. 
By default, we assume that the content is fetched via the path with the highest available transmission rate. 
The requesting vehicle can access the content in three ways: 1) popular content is accessed from local RSUs, 2) popular content is accessed from neighboring RSUs, and 3) popular content is accessed from the base station, and each of them has different characteristics; the total transmission delay is also different.

When $V_i$ sends a request, if this RSU has already cached the content. According to Shannon's channel capacity theorem, the theoretical maximum transmission rate between vehicle $V_i$ and RSU $R_j$ over an additive white Gaussian noise (AWGN) channel can be expressed as:

\begin{equation}
    r_{i,j}=W_{i,j}log_2(1+\frac{P_id_{i,j}^{-\beta}h_{i,j}^2}{\sigma_{i,j}^2}),
\end{equation}
where $W_{i,j}$ and $h_{i,j}$ are denoted as the bandwidth and fading factor of the vehicle $V_i$ and the RSU $R_j$, $d_{i,j}$ is the distance between $V_i$ and $R_j$, where $\beta$ is the path loss factor, $\sigma^2_{i,j}$ and $P_i\in{[0,P_{max}]}$ denote the data upload and Gaussian white noise power.

It should be noted that $r_{i,j}$ denote the upper bound of the theoretical transmission rate, which assumes ideal channel conditions including perfect channel state information and optimal coding schemes. This provides a performance benchmark for evaluating our caching strategy, consistent with standard approaches in vehicle-to-everything communication research\cite{9399641,10285090,9139263,9944845}.

Correspondingly, its transmission time $t_{i,j}$ is

\begin{equation}
    t_{i,j}=\frac{d_i}{r_{ij}}.
\end{equation}

If the content requested by the vehicle is not in the local RSU $R_j$ but is stored in the RSU $R_k$ within its range, the vehicle needs to obtain the content from the RSU $R_k$ with transmission rate $r_{i,j}$ and transmission time $t_{i,j}$, respectively:

\begin{equation}
    r_{i,k}=W_{i,k}log_2(1+\frac{P_Rd_{i,k}^{-\beta}h_{i,k}^2}{\sigma_{i,k}^2}),
\end{equation}

\begin{equation}
    t_{i,k}=\frac{d_i}{r_{i,j}}+\frac{d_i}{r_{i,k}}.
\end{equation}

If the content requested by the vehicle is neither in the local RSU $R_j$ nor in the neighboring RSU $R_k$, the vehicle needs to obtain the content from the BS with the transmission rate $r_{i,B}$ and the transmission time $t_{i,B}$, respectively:

\begin{equation}
    r_{i,B}=W_{i,B}log_2(1+\frac{P_Dd_{i,B}^{-\beta}h_{i,B}^2}{\sigma_{i,B}^2}),
\end{equation}

\begin{equation}
    t_{i.B}=\frac{d_i}{r_{i,B}}.
\end{equation}

\section{COOPERATIVE CACHING SCHEME}\label{se:4}
In this section, we detail our technical solution. We first protect user information through an asynchronous federated learning algorithm, and then we propose a GRU-VAE based algorithm to predict the popular content in the next time slot of the vehicle, and then the predicted data of the popular content in the next moment is inputted into a SAC-based decision model to determine the optimal caching location.
The rationale for content popularity prediction is detailed in the supplementary materials.

\begin{figure}
    \centering
    \includegraphics[width=1\linewidth]{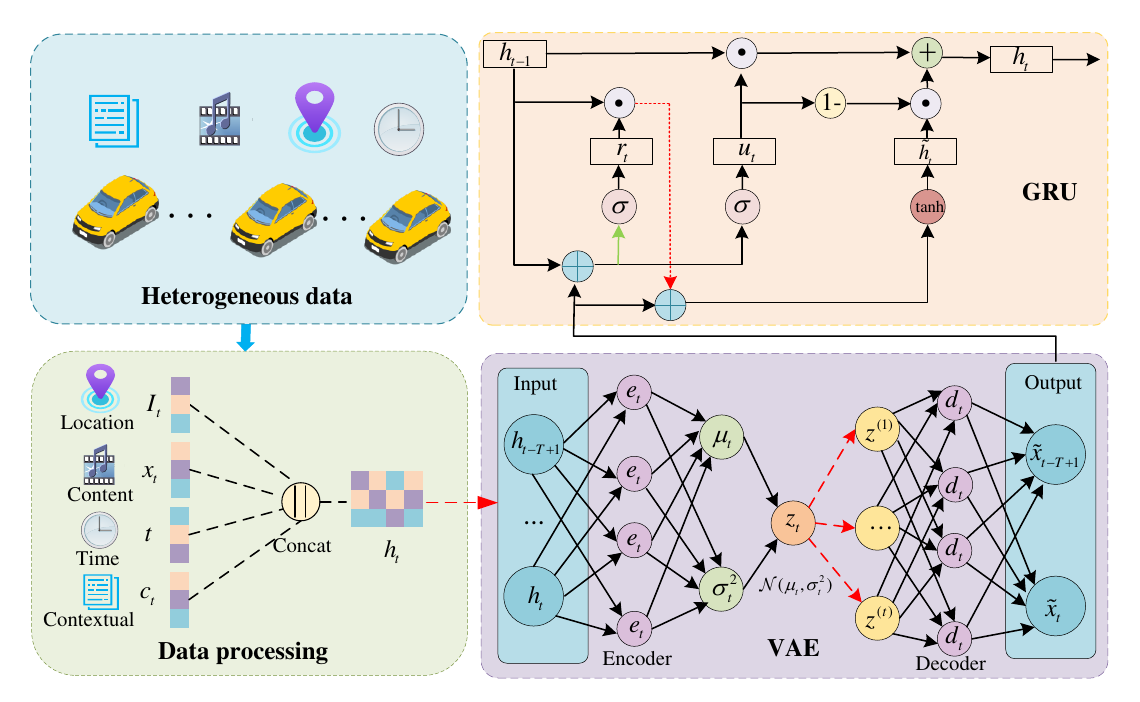}
    \caption{GRU-based VAE framework.}
    \label{fig_gru}
\end{figure}

\subsection{Content Popularity Prediction}\label{se:4a}
As shown in Fig.~\ref{fig_gru}, the content prediction model is divided into the following four main steps.

1) Data preprocessing: In the vehicle dynamic data input stage, it is first necessary to integrate heterogeneous data from multiple sources, including user request content $x_t$ (e.g., video ID, data size), vehicle location 
$I_t$, timestamps $t$, and contextual features $c_t$ (e.g., traffic events). Position and time are mapped into low-dimensional continuous vectors through the embedding layer and concatenated with high-dimensional content features via Eq.~\eqref{eq:11} to form a unified input vector $h_t$,which provides structured, low-noise inputs for subsequent models while capturing spatiotemporal dynamics.

\begin{equation}
    h_t=Concat(x_t,Embedding(I_t),Embedding(t),c_t).
    \label{eq:11}
\end{equation}

2) Latent feature extraction for VAE: The core of this step is to extract semantic features of the content (e.g., user preferences, topic relevance) and to address data sparsity and noise. Variational Auto-Encoder (VAE) maps the high-dimensional input feature $h_t$ to the low-dimensional latent space through an encoder, generating latent variables whose mean $\mu_t$ and variance $\sigma^2$ of the distribution in the latent space obey a Gaussian distribution.

\begin{equation}
    \mu_t=f_{enc}(h_t;W_{enc}), \sigma_t=g_{enc}(h_t;W_{enc}),
\end{equation}
where $f_{enc}$ and $g_{enc}$ are the neural networks of the encoder and $W_{enc}$ is the weight parameter of the encoder.

The encoder then samples the latent variable $z_t$ from a Gaussian distribution using a reparameterization trick:

\begin{equation}
    z_t=\mu_t+\sigma\odot\varepsilon, \quad\varepsilon\sim\mathcal{N}(0,I),
\end{equation}  
where $\varepsilon$ is the noise sampled from the standard normal distribution.

The decoder then maps the resulting latent variable $z_t$ back into the data space, reconstructing the original input $\hat{x}$, the
\begin{equation}
    \hat{x_t}=f_{dec}(z_t; W_{dec}),
\end{equation}
where $f_{dec}$ is the neural network of the decoder and $W_{dec}$ is the weight parameter of the decoder.

We use the mean square error (MSE) to measure the difference between the input data and the reconstructed data and add a Kullback-Leibler (KL) divergence regularization term to measure the difference between the latent distribution $q(z_t|x_t)$ and the prior distribution $p(z_t)$, so that the loss function of the VAE can be expressed by the following equation:
\begin{equation}
    \mathcal{L}_{VAE}=\mathbb{E}_{x_t}[\|x_t-\hat{x}_t\|^2]+\beta\cdot KL(q(z_t|x_t)\|p(z_t)),
\end{equation}
where $KL(q(z_t|x_t)\|p(z_t))=\frac{1}{2}({\displaystyle\sum^d_{i=1}(\mu_i^2 + \sigma_i^2 - log\sigma_i^2-1)}$, and $d$ is the dimensionality of the latent space.

3) Timing modeling of GRU: The gated recurrent unit (GRU) receives the sequence of latent variables $\hat{x_t}$ output by the VAE and captures temporal dependencies through update gates and reset gates. The update gate $u_t$ controls how much historical information $h_{t-1}$ is retained, while the reset gate $r_t$ determines how much of $h_{t-1}$ is forgotten. They are defined as
\begin{equation}
    u_t=\sigma(W_u\cdot[h_{t-1},\hat{x_t}]+b_u),
\end{equation}
\begin{equation}
    r_t=\sigma(W_r\cdot[h_{t-1},\hat{x_t}]+b_r),
\end{equation}
where $W_u,W_r$ and $b_u,b_r$ denote the weight parameters and bias terms of the update and reset gates, respectively, $[h_{t-1},\hat{x_t}]$ is the concatenation of the previous hidden state and the current input, and $\sigma$ is the sigmoid activation function.

The candidate state $\tilde{h}_t$ is then obtained from the combination of the current inputs and the gating via tanh activation function with the following mathematical expression:
\begin{equation}
    \tilde{h}_t=tanh(W_h\cdot[r_t\odot h_{t-1},\hat{x_t}]+b_h),
\end{equation}
where $W_h$ and $b_h$ are the weight parameters and bias terms of the candidate hidden states, the final hidden state $h_t$ combines the weighted combination of the hidden state and the candidate hidden state at the previous moment by updating the gate, and then the final hidden state $h_t$ predicts the content popularity at the next moment $y_t$ through the fully connected layer with the following mathematical expression:
\begin{equation}
  h_t=(1-u_t)\odot h_{t-1}+u_t\odot \tilde{h}_t,  
\end{equation}
\begin{equation}
    \hat{y}_t=Softmax(W_t\cdot h_t+b_o),
\end{equation}
where $W_t$ and $b_0$ are the weight parameters and bias terms of the fully connected layer, and we use the MSE to measure the gap between the prediction and the real request $y_t$, so the loss function of the GRU can be expressed by the following equation:
\begin{equation}
    \mathcal{L}_{GRU}=\mathbb{E}_{t}[\|y_t-\hat{y}_t\|^2].
\end{equation}

4) Joint training and optimization: the model adopts a staged training strategy: first pre-train the VAE to learn the robust latent space, subsequently freeze the VAE parameters to train the GRU, and finally jointly fine-tune the overall model. The joint loss function $\mathcal{L}_{total}$ consists of the loss of the VAE weighted by the loss of the MSE of the GRU, with the following formula:
\begin{equation}
 \mathcal{L}_{total}=\lambda\mathcal{L}_{VAE}+(1-\lambda)\mathcal{L}_{GRU},
\end{equation}
where $\lambda$ is a hyperparameter that regulates the relative importance of VAE and GRU losses. For further details on the relevant theory, please refer to the supplementary materials.

\subsection{Asynchronous Federated Learning}\label{se:4b}

Given that the mobility of vehicles results in a dynamically changing communication environment and data distribution state, asynchronous federated learning better accommodates this characteristic by permitting nodes to upload local model updates in a time-sharing manner. As illustrated in Fig. \ref{fig_fl}, the core steps of our designed vehicle asynchronous federated learning framework are as follows:

\begin{figure}
    \centering
    \includegraphics[width=1\linewidth]{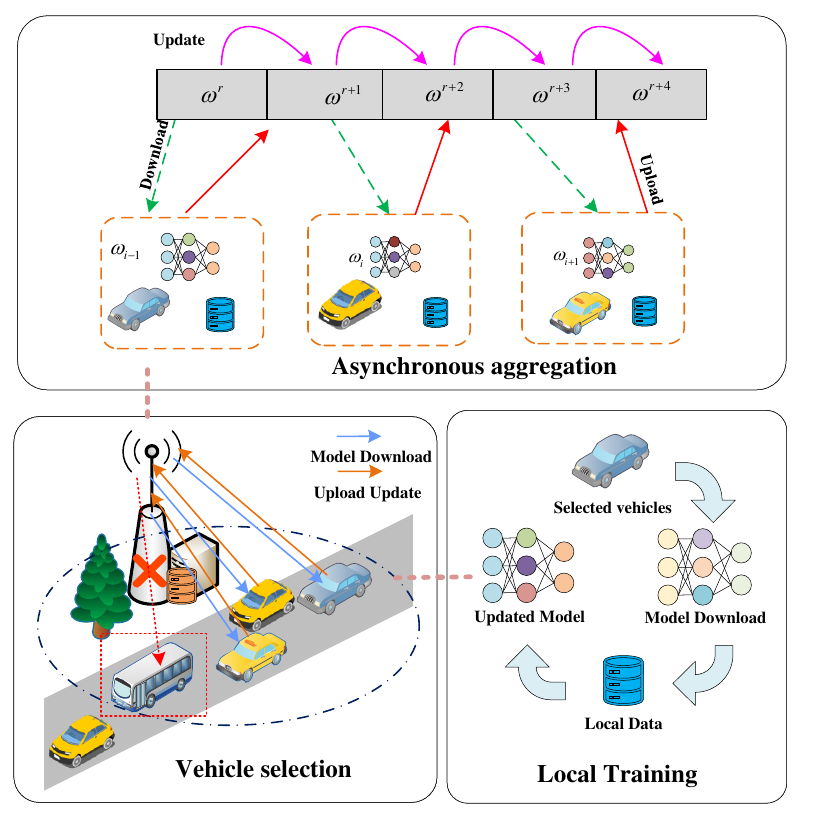}
    \caption{Asynchronous FL.}
    \label{fig_fl}
\end{figure}

1) Vehicle selection: DT predicts the dwell time of vehicles within the RSU coverage area by monitoring their mobility status in real time. This predictive information is relayed to the AFL module to inform vehicle selection. Based on the dwell time predictions provided by DT, AFL selects vehicles whose dwell time exceeds both its local training duration and model upload time, thereby ensuring the selected vehicles can fully participate in the current training cycle.

2) Local training: DT synchronises vehicle status data in real time, providing AFL with the contextual information required for local training. This information assists AFL in understanding each vehicle's data characteristics and training environment, thereby enhancing local model training. AFL utilises the contextual information provided by DT, combining it with the global model $w^r$ and local data to train the local model. By incorporating a regularisation term into the objective function $J(\omega_{i,j})$, AFL constrains the deviation between the local model and the global model, ensuring the stability and consistency of local training. The specific objective function is:

\begin{equation}
    J(\omega_{i,j}^r)=f(\omega_{i,j}^r)+\frac{\kappa}{2}\|\omega_{i,j}^r-\omega^r\|^2,
\end{equation}
where $\kappa$ is the regularization coefficient, $j$ is the number of iterations, $f(\omega_{i,j}^r)=\frac{1}{K}\sum^K_{K=1}\mathcal{L}_{total}^i(\omega_{i,j}^r)$, and K is the number of data. To update the local model parameters $\omega_{i,j}^r$, we compute the gradient of the objective function $J(\omega_{i,j}^r)$, which consists of two parts: the gradient of the local loss function and the regularization term. Thus, the gradient is given by

\begin{equation}
    \bigtriangledown J(\omega_{i,j}^r)=\eta\bigtriangledown f(\omega_{i,j}^r)+\eta\kappa(\omega_{i,j}^r-\omega),
\end{equation}
where $\eta$ is the learning rate. Then the local model $\omega_{i,j+1}^r$ can be obtained for the next round.

3) Model Upload: After the vehicle completes local training, the trained model $\omega_j$ is uploaded to the local RSU.

4) Asynchronous Aggregation: DT provides real-time updates on vehicle data quality and location information through its global heatmap and virtual mirror modules. This information assists RSUs in more accurately evaluating each vehicle's contribution during asynchronous aggregation. AFL dynamically adjusts the weight distribution within the aggregation formula using the data quality and location weights $\alpha_1$ and $\alpha_2$ provided by DT. The specific aggregation formula is

\begin{equation}
    \omega^{r+1}=\omega^r+\varrho\omega_j,
\end{equation}
where $\varrho=\alpha_1\frac{n_k}{\sum_{i\in S}n_i}+\alpha_2\frac{L_i^r}{L_s},$, $n_k$ is the volume of data for vehicle $V_k$, $\alpha_1$ and $\alpha_2$ are the volume weights and location weights, and there is $\alpha_1+\alpha_2=1,\alpha_1,\alpha_2\in[0,1]$. This weighting mechanism effectively distinguishes between vehicles with “large amounts of low-quality data” and those with “small amounts of high-quality data” by comprehensively considering both data quantity and location information.

5) Global update: DT provides AFL with contextual information for global model updates by continuously monitoring network conditions and vehicle distribution. Based on this contextual information supplied by DT, AFL broadcasts the updated global model $\omega^{r+1}$ to all vehicles. Concurrently, AFL dynamically adjusts the strategy for the next training round using DT's feedback to adapt to changes in network conditions. The specific training procedure is described in Algorithm~\ref{alg:vfl}.

\begin{algorithm}
\caption{Vehicular Asynchronous Federated Learning (VAFL)}
\label{alg:vfl}
\begin{algorithmic}[1]
\REQUIRE Initial model $\omega^0$, learning rate $\eta$, regularization $\kappa$, data weight $\alpha_1$, location weight $\alpha_2$, RSU coverage $L_s$, total rounds $R$
\ENSURE Optimized model $\omega^R$

\STATE \textbf{Server Execution:}
\STATE Initialize global model $\omega^0$
\FOR{$r=0$ to $R-1$}
    \STATE Select vehicles $S_r$ with $T_{\text{stay}}^i > T_{\text{train}}^i + T_{\text{trans}}^i$
    \STATE Compute $\text{sum\_n} \gets \sum_{i \in S_r} n_i$
    
    \FORALL{vehicle $V_i \in S_r$ \COMMENT{Parallel processing}}
        \STATE Download $\omega^r$ to $V_i$
        \STATE $\omega_i \gets$ ClientUpdate($V_i, \omega^r$)
        \STATE Upload $\omega_i$ to RSU
        \STATE $\varrho_i \gets \alpha_1\frac{n_i}{\sum_{i\in S}n_i} + \alpha_2\frac{L_i^r}{L_s}$
        \STATE $\omega^{r+1} \gets \omega^r + \varrho_i\omega_i$ \COMMENT{Async aggregation}
        \STATE Broadcast $\omega^{r+1}$
    \ENDFOR
\ENDFOR

\STATE \textbf{ClientUpdate($V_i, \omega^r$):}
\STATE Initialize $\omega_{i,0}^r \gets \omega^r$
\FOR{$j=0$ to $J-1$}
    \STATE Compute gradient $\nabla J = \nabla f(\omega_{i,j}^r) + \eta\kappa(\omega_{i,j}^r - \omega^r)$
    \STATE Update $\omega_{i,j+1}^r \gets \omega_{i,j}^r - \nabla f(\omega_{i,j}^r) - \eta\kappa(\omega_{i,j}^r - \omega^r)$
\ENDFOR
\RETURN $\omega_{i,J}^r$
\end{algorithmic}
\end{algorithm}

\subsection{Cache Allocation Scheme Based on DRL}

After the local RSU collects predicted popular content from vehicles in the region, it needs to make edge cache allocation decisions on this content data to minimize the content delivery delay and improve the cache hit ratio. Thus, we model the cache decision problem as an MDP problem\cite{10379478}, and we refine the underlying SAC to solve the problem. The state space can be represented as $S=\{c_1,c_2,\cdots,c_n,\phi_1,\phi_2,\cdots,\phi_m\}$, $c_i$ includes the size, ID and access times of the cached content $i$, and $\phi_j$ denotes the characteristics of request $j$. The action space can be represented as $A=\{a_{1}(t),a_2(t),\cdots,a_m(t)\}$, where $a_i$ denotes the caching decision of the content $i$ at the time slot $t$, and the local RSU will replace the content $i$ if $a_i=1$, and will not replace the content if $a_i=0$.

The objective of this paper is to achieve optimal cache utility, the reward function $R$ should react to the effect of cache allocation, and its design objectives include: reduce content delivery delay and improve cache hit ratio, so the vehicle $V_i$ in time slot $t$ request content $f$ reward function can be designed as the following form:

\begin{equation}
    \begin{split}
    R_i^f(t)=  \left\{ \begin{array}{rcl}
 -\lambda_1(\xi t_{i,j}-\zeta(1-\frac{\lambda_1\sum_{f\in j}N_i}{\sum_iQ_i})),&  \mbox{local}\\ 
 -\lambda_2(\xi t_{i,k}-\zeta(1-\frac{\lambda_2\sum_{f\in k}N_i}{\sum_iQ_i})),&  \mbox{R-R,}\\
 -\lambda_3(\xi t_{i,B}-\zeta(1-\frac{\lambda_3\sum_{f\in B}N_i}{\sum_iQ_i})),& \mbox{BS}
 \end{array}\right.
\end{split}
\end{equation}
where $\lambda_1+\lambda_2+\lambda_3=1$ and $0<\lambda_1<\lambda_2\ll\lambda_3$,$\xi+\zeta=1,\xi,\zeta\in(0,1)$, and assuming that $\delta_i$ is the contents of $V_i$ requests quantity,therefore, the reward function $R(t)$ can be calculated by the following equation:

\begin{equation}
    R(t)=\sum_{i=1}^K\sum_{f=1}^{\delta_i}R_i^f(t).\label{eq:28}
\end{equation}

\begin{figure*}[t]
    \centering
    \includegraphics[width=1\linewidth]{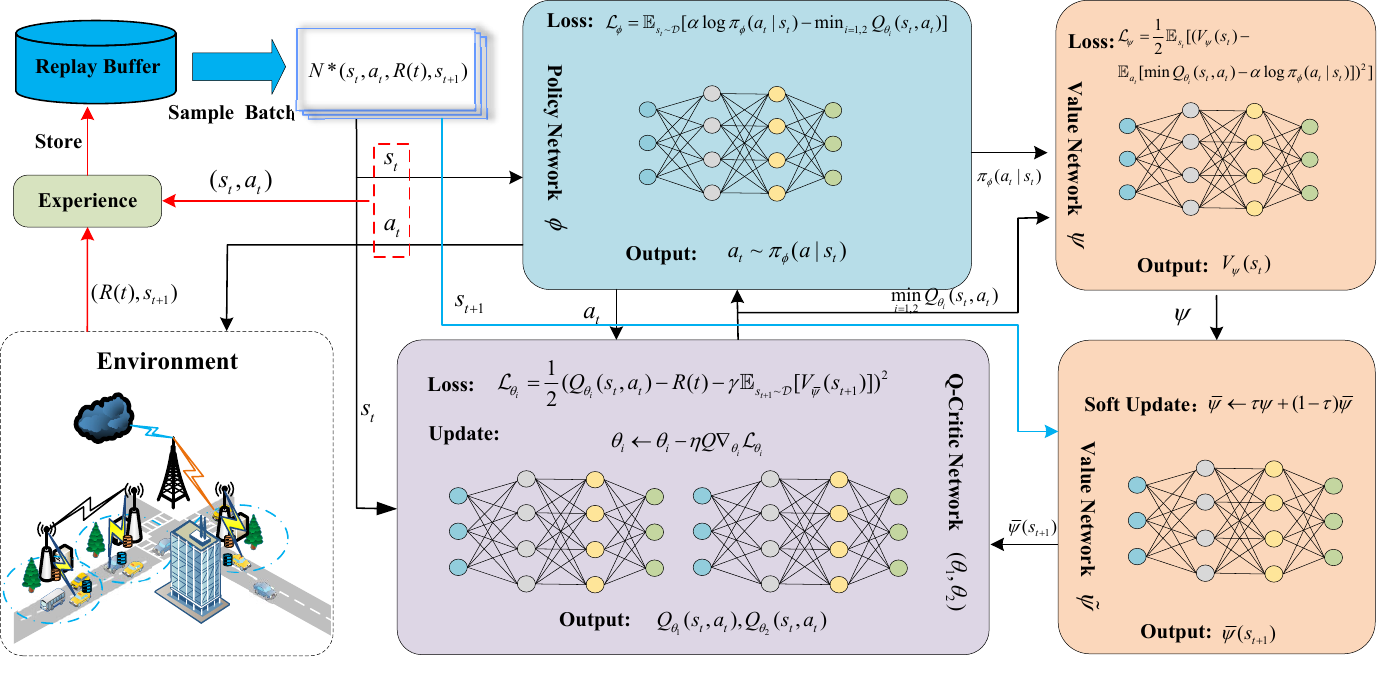}
    \caption{SAC algorithm structure diagram.}
    \label{fig:enter-label}
\end{figure*}

As shown in Fig.~\ref{fig:enter-label}, the SAC network structure consists of six main modules: the strategy (actor) network $\phi$, the value-criticism network $\psi$, the value-targeting network $\bar{\psi}$, the Q-criticism network $\theta_1$,the Q-targeting network $\theta_2$, the replay buffer and the environment. The strategy network $\pi_{\phi}(a_t|s_t)$ outputs probability distributions of actions and is used to generate cache allocation decisions with the goal of maximizing the expected cumulative reward and strategy entropy. For each time step $t$, the intelligent body samples a random batch of experiences $(s_t,a_t,R(t),s_{t+1})$ from the experience playback buffer $\mathcal{D}$, and then based on the maximum entropy reinforcement learning and Bellman's equations, the state value of the current state $s_i$ can be estimated

\begin{equation}
   \begin{split}
           V(s_t)=&\min_{i=1,2}\mathbb{E}_{s_t\sim \mathcal{D},a_t\sim \pi_\theta}[Q_{\theta_i}(s_t,a_t)\\
           &-\alpha \log\pi_\phi(a_t|s_t)],
   \end{split}
\end{equation}
where $\alpha$ is the entropy regularization factor, and $\log\pi_\phi(a_t|s_t)$ is the entropy of the strategy $\pi_\theta$. The parameters $\psi$ of the value network are then updated using the loss function (\ref{eq:29})

\begin{equation}
    \mathcal{L}_{\psi}=\frac{1}{2}(V_\psi(s_t)-Q_\theta(s_t,a_t)+\log\pi_\phi(a_t|s_t))^2.
    \label{eq:29}
\end{equation}

According to the soft Bellman equation, an estimate of the $(s_t,a_t)$ action state value can be obtained:

\begin{equation}
    Q_{target}(s_t,a_t)=R(t)+\gamma\min_{j=1,2}\bar{\psi}_j(s_{t+1}),
\end{equation}
where $\bar{\psi_j}(s_{t+1})$ is the output of the target-value network, so the loss of the Soft-Q network can be calculated by the following equation:

\begin{equation}
    \mathcal{L}_{\theta_i}=\mathbb{E}_{(s_t,a_t)}[(Q_{\theta_i}(s_t,a_t)-Q_{target}(s_t,a_t))^2].
    \label{eq:31}
\end{equation}

Then, the policy network parameters $\phi$ are updated to maximize the entropy and Q values of the policy:

\begin{equation}
    \mathcal{L}_\phi=\mathbb{E}_{s_t\sim\mathcal{D}}[\alpha\log\pi_\phi(a_t|s_t)-\min_{i=1,2}Q_{\theta_i}(s_t,a_t)].
\end{equation}

Finally, use the soft update method to update the target Critic network parameters:
\begin{equation}
    \bar{\psi}_i=\tau\psi_i+(1-\tau)\bar{\psi}_i,
\end{equation}
where $\tau$ is the update coefficient, we use the SAC-based cache allocation model for the training process see Algorithm~\ref{algo_SAC}.

\begin{algorithm}
    \caption{Cooperative Caching Allocation Based on SAC Algorithm}
    \label{algo_SAC}
    \begin{algorithmic}[1]
        \STATE \textbf{Input:} State space $\mathcal{S}$,Replay memory $\mathcal{D}$, discount factor $\gamma$
        \STATE \textbf{Output:}Networks $\phi$, $\psi$, $\bar{\psi}$, $\theta_i,i=1,2$
        \STATE Initialize policy network \( \phi \), Q-networks \( Q_{\theta_1}, Q_{\theta_2} \), value network \( V_\psi \)
        \STATE Initialize replay buffer \( \mathcal{D} \)

        \FOR{each episode}
            \STATE Initialize state \( s_t \)
            \FOR{each time step \( t \)}
                \STATE Sample action \( a_t \sim \pi_\phi(a_t|s_t) \)
                \STATE Execute action and observe reward \( R(t) \) and next state \( s_{t+1} \)
                \STATE Store transition \( (s_t, a_t, R(t), s_{t+1}) \) in \( \mathcal{D} \)

                \IF{size of \( \mathcal{D} \) $>$ batch size}
                    \STATE sample a batch from \( \mathcal{D} \)

                    \STATE Update value network $\psi$ by eq.~\ref{eq:29}
                    \STATE Update soft-Q networks $\theta_i$ by eq.~\ref{eq:31}
                    \STATE Update policy network $\phi$ by
                    \[\mathcal{L}_{\phi} = \mathbb{E}_{s_j \sim \mathcal{D}}[\alpha \log \pi_\phi(a_j|s_j) - \min_{i=1,2} Q_{\theta_i}(s_j, a_j)]\]

                    \STATE Update target value networks $\bar{\psi}$ by
                    \[
                    \bar{\psi}\leftarrow\tau \psi + (1 - \tau) \bar{\psi}
                    \]

                \ENDIF
            \ENDFOR
        \ENDFOR
    \end{algorithmic}
\end{algorithm}

\section{Simulation Results and Analysis}

\subsection{Parameter Settings}

\begin{figure}[!htbp]
    \centering
    \begin{minipage}{0.47\linewidth}
        \centering
        \includegraphics[width=\linewidth]{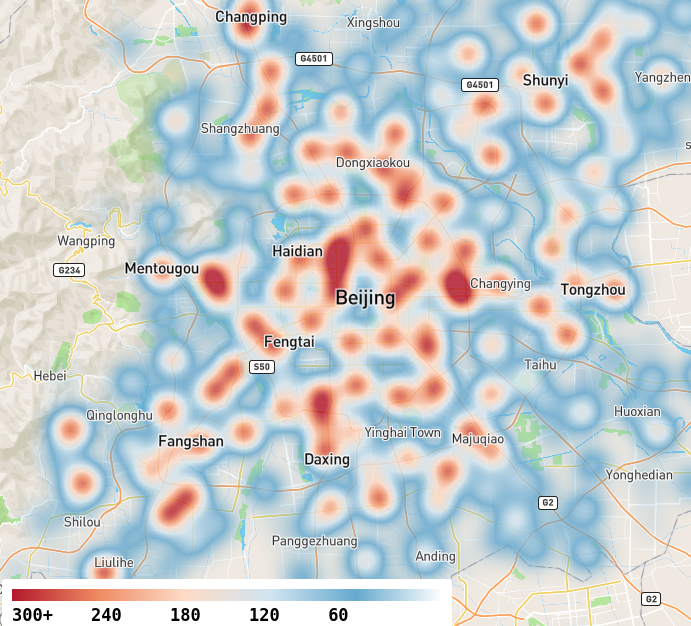}
        \footnotesize{(a) 7:00-9:00}
        \vspace{4px}
        \label{fig_a}
    \end{minipage}
    \hfill  
    \begin{minipage}{0.47\linewidth}
        \centering
        \includegraphics[width=\linewidth]{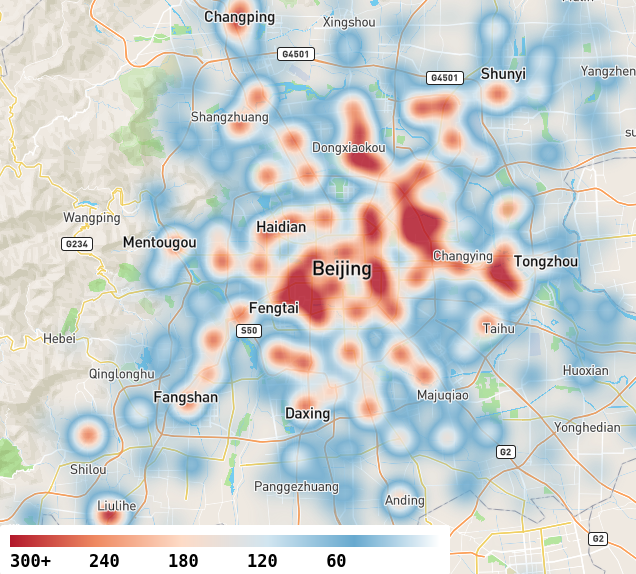}
        \footnotesize{(b) 11:00-13:00}
        \label{fig_b}
    \end{minipage}
    
    \begin{minipage}{0.47\linewidth}
        \centering
        \includegraphics[width=\linewidth]{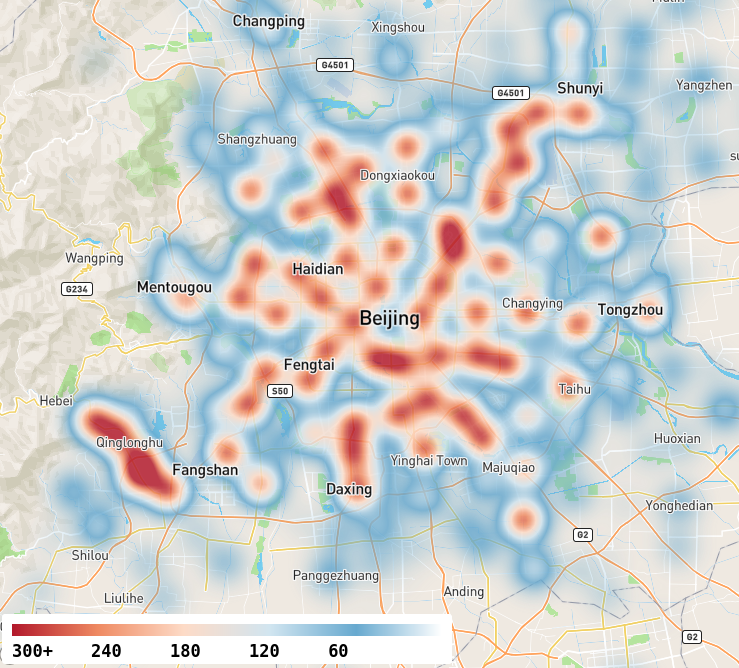}
        \footnotesize{(c) 17:00-19:00}
        \label{fig_c}
    \end{minipage}
    \hfill
    \begin{minipage}{0.47\linewidth}
        \centering
        \includegraphics[width=\linewidth]{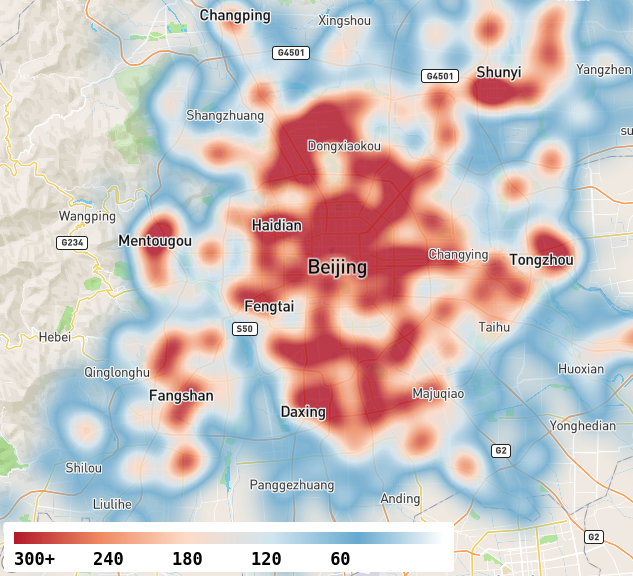}
        \footnotesize{(d) 23:00-01:00}
        \label{fig_d}
    \end{minipage}
    \caption{Heat map of the distribution of vehicles under different scenarios.}
    \label{fig_}
\end{figure}

\begin{figure}[!htbp]
    \centering
    \begin{minipage}{0.47\linewidth}
        \centering
        \includegraphics[width=\linewidth]{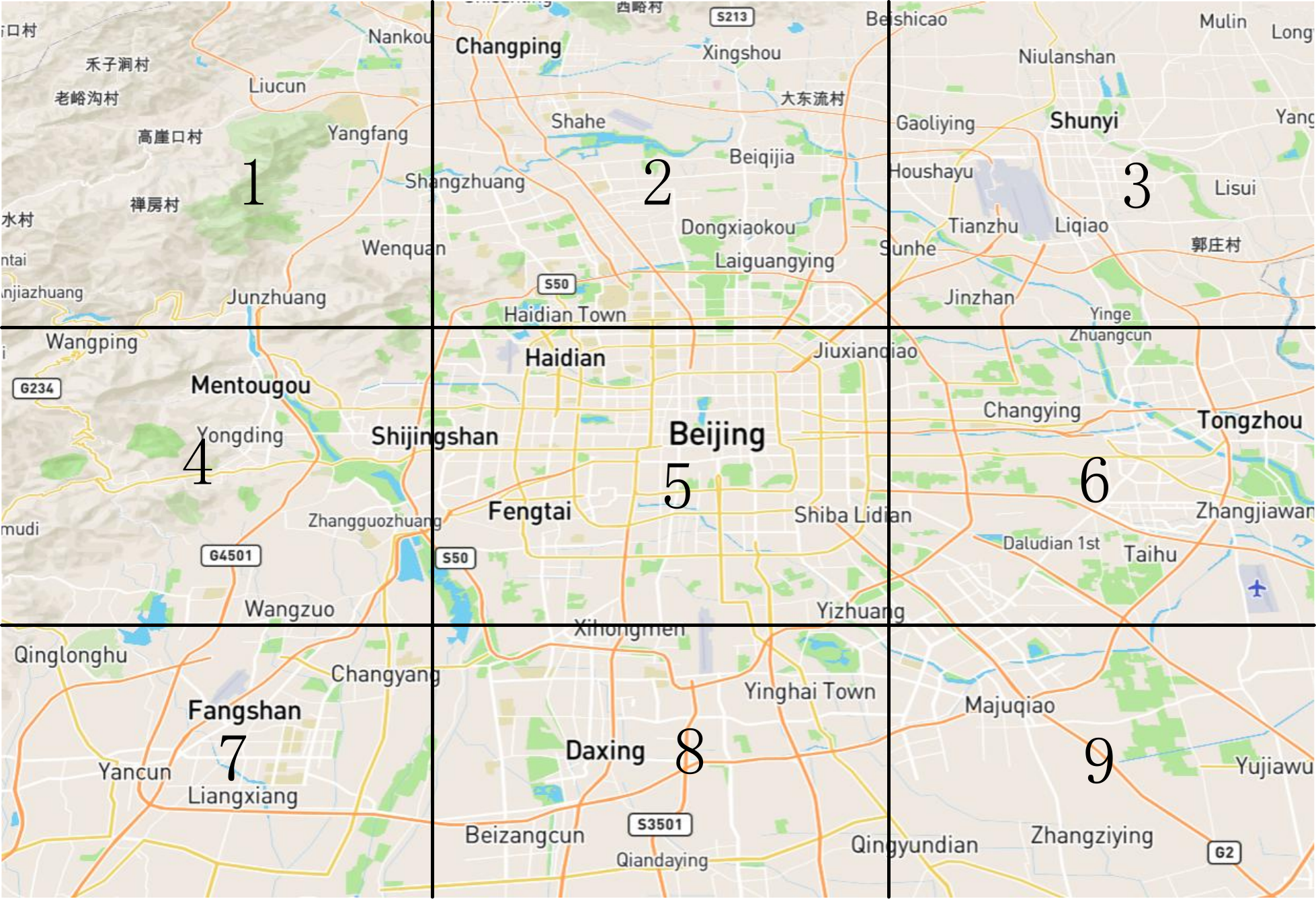}\vspace{8px}
        \footnotesize{(a) Schematic of urban area division.}
        \label{fig__a}
    \end{minipage}
    \begin{minipage}{0.47\linewidth}
        \centering
        \includegraphics[width=\linewidth]{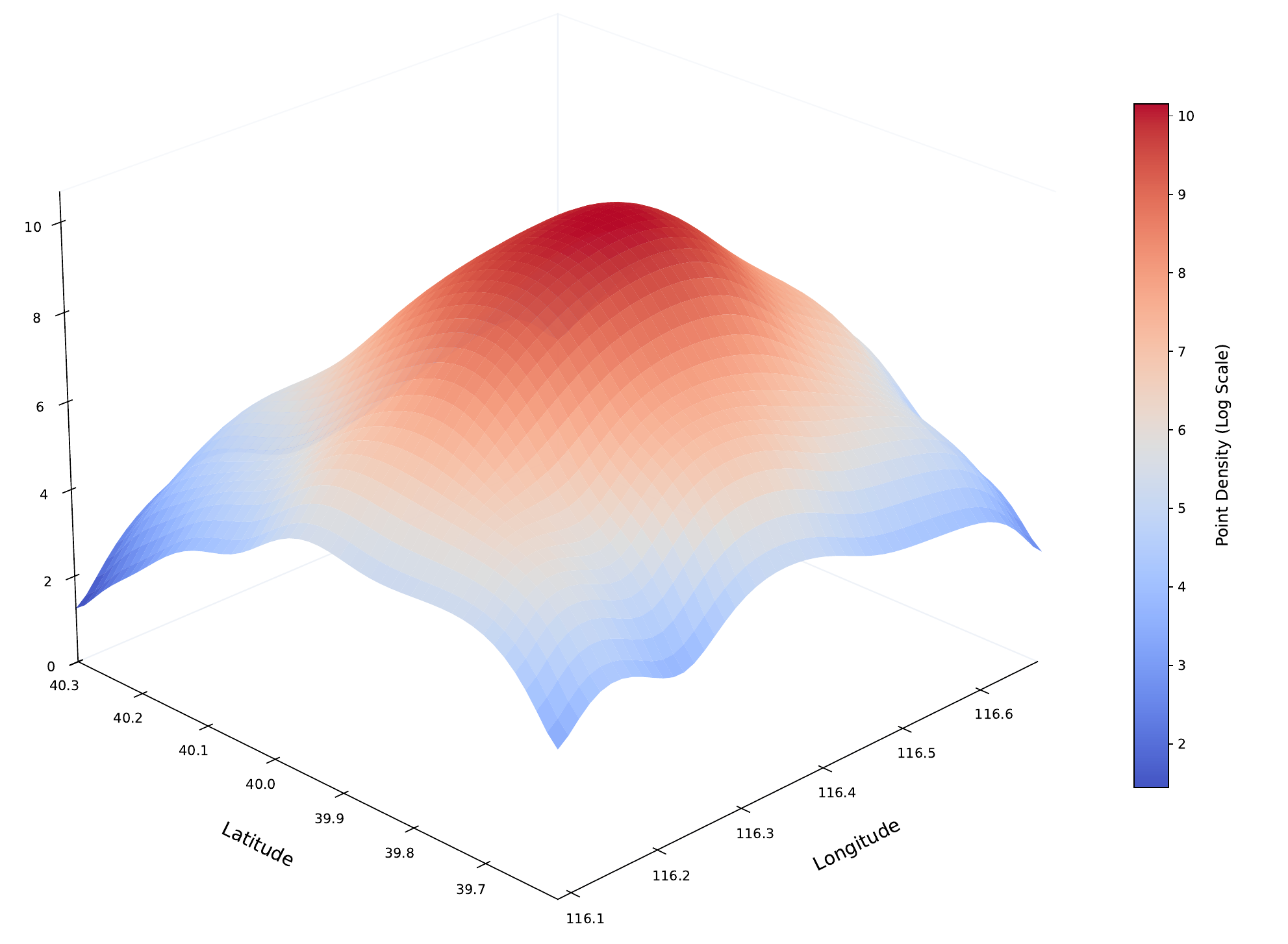}
        \footnotesize{(b) Vehicle 3D heatmap.}
        \label{fig__b}
    \end{minipage}
    \caption{Area division and 3D heatmap.}
    \label{fig__}
\end{figure}

In this section, we evaluate the performance of the provided DAPR scheme. In the modeling experiments, to be closer to real scenarios, we choose to use real data for evaluation. We respectively use the T-Drive trajectory data sample\footnote{https://www.microsoft.com/en-us/research/publication/t-drive-trajectory-data-sample/}\cite{10.1145/2020408.2020462,10.1145/1869790.1869807}, Top 5000 Albums of All Time Dataset\footnote{https://www.kaggle.com/datasets/michaelbryantds/top-5000-albums-of-all-time-rateyourmusiccom/data}, YouTube data set\footnote{https://www.kaggle.com/datasets/nnqkfdjq/statistics-observation-of-random-youtube-video/data} and MovieLens 1M\footnote{https://grouplens.org/datasets/movielens/1m/} as vehicle data and content data. Fig.~\ref{fig_} shows the heat map of vehicle trajectories in Beijing at different times. In order to measure the variations and differences in the number of vehicles in different regions, as shown in Fig.~\ref{fig__}, we divide Beijing into nine square regions. We assume that there are a number of servers in each region, and we treat each region as a whole from the perspective of a service provider. If a more fine-grained allocation scheme is needed, simply increase the number of divided regions. The dataset contains the Global Positioning System (GPS) trajectories of 10,357 cabs in Beijing between February 2 and February 8, 2008, which contains about 15 million GPS points in total, and the total mileage of the trajectories reaches 9 million kilometers. In preprocessing the vehicle data, we define the speed in each vehicle data record as the average speed calculated by dividing distance by time. With respect to speed and time, we take the entire dataset and calculate the number of vehicles in each area during each time interval, i.e., the vehicle density in each area. Then, for each time interval, we summed the speeds of all vehicles in each area and divided by the total number of vehicles to calculate the traffic speed in that area for that time interval. 

MovieLens 1M includes 1000209 ratings from 6040 anonymous mobile users for approximately 3800 movies, with ratings ranging from one to five stars, and the dataset includes user information (gender, age, occupation, zip code), movie information (title, genre), and ratings records. 
The YouTube dataset contains about 1,500 videos released in April 2018, with hourly observations over the last month (views, comments, likes, etc.).
 The dataset contains the top 5k albums of all time as selected by users of the rateyourmusic.com community. The Top 5000 Albums of All Time Dataset was acquired via crawling on October 12, 2021, and includes the rankings, album name, artist name, release date, genre, description, average rating, number of ratings, and comment number of these attributes. The vehicles randomly select 98\% of the data from the local data as the training set and 2\% of the data as the test set. In each round, each vehicle randomly selects a portion of content from the test set as its requested content. Table~\ref{tab_2} lists the other parameters of the associated task and environment.

\begin{table}[htbp]
\caption{Environment Parameters}
\begin{center}
\begin{tabular}{@{}p{6cm}p{2cm}@{}}
\toprule
\textbf{Parameters of System Model} & \textbf{Value} \\ \midrule
free-flow velocity $v_f$  & 60 km/h \\
Maximum density $\rho_{max}$ & 100 vehicles/km\\
bandwidths $W_{i,j},W_{i,k},W_{i,B}$ & 540 kHz\\
Additive white Gaussian noise$\sigma^2$ & -114dBm\\
RSU coverage length & 1000 m \\
Vehicle transmission power $P_i$ & 3 dBm\\
RSU transmission power $P_R$ & 30 dBm\\
BS transmission power $P_B$ & 43 dBm\\
Optimizer  & Adam\\
Replay buffer size $\mathcal{D}$ & 100000\\
The learning rate of actor network & $10^{-4}$\\
The learning rate of critic network & $10^{-4}$\\
The discount factor $\gamma$ & 0.99\\
The soft update factor $\tau$ & 0.1\\
The batch size of SAC & 64 \\
The learning rate in the GRU & 0.001\\
The learning rate in the VAE & 0.001\\
\bottomrule
\end{tabular}
\label{tab_2}
\end{center}
\end{table}

\subsection{Performance Metrics Explanation}

To ensure reproducible and interpretable results, we explicitly define the evaluation metrics:

\begin{itemize}
    \item Cumulative Reward: Represents the total benefit gained by the caching strategy throughout the entire testing cycle. Calculated using Eq. (\ref{eq:28}), this is a composite utility metric that comprehensively reflects the combined benefits of reduced response times, bandwidth savings, and enhanced user experience resulting from cache hits. A higher reward value indicates that the algorithm can make more optimal caching decisions during long-term operation.
    \item Cache Hit Ratio: The percentage of requests that can be served directly from the cache. This metric directly reflects cache efficiency; a higher hit rate means fewer requests need to access the backend server.
    \item Transmission Delay: End-to-end latency from request to content delivery. Includes V2R, R-R, and V2B link delays.
    \item Prediction Loss (MSE): MSE measures the deviation between the predicted popularity of content and its actual popularity.
\end{itemize}

\subsection{Performance Comparison}\label{Se:5}

To evaluate the effectiveness of our scheme, we compare the cache scheduling algorithms under the same digital twin-enabled vehicle edge caching environment:

\begin{itemize}
    \item $\epsilon$-greedy: Selects the most requested content with probability 1-$\epsilon$ and randomly selects content with probability $\epsilon$ for caching at local RSU. In simulation, $\epsilon$=0.1.
    \item CMCF\cite{10173652}: Predicts content popularity using auto-encoder and optimizes cache allocation with DDQN, proactively caching potentially popular content to improve resource utilization.
    \item SDDPG\cite{9234632}: Uses GRU to predict popular content and combines DDPG algorithm for cache allocation, learning optimal policies to adapt to dynamic request patterns.
    \item CAFR\cite{9944845}: Predicts content popularity using auto-encoder and applies Dueling-DQN for cache allocation. By decomposing the value function into state-value and advantage, it evaluates different cache decisions more accurately.
    \item DTSO2C\cite{WANG2025103681}: A DDQN-based two-stage online computation offloading and application caching algorithm that employs a dual-timescale framework with Lyapunov optimization to stabilize cache update costs. It makes offloading decisions per time slot and caching decisions per time frame, adapting to highly dynamic service popularity.
    
\end{itemize}


\begin{figure*}[htbp]
    \centering
    \begin{minipage}{0.32\linewidth}
        \centering
        \includegraphics[width=\linewidth]{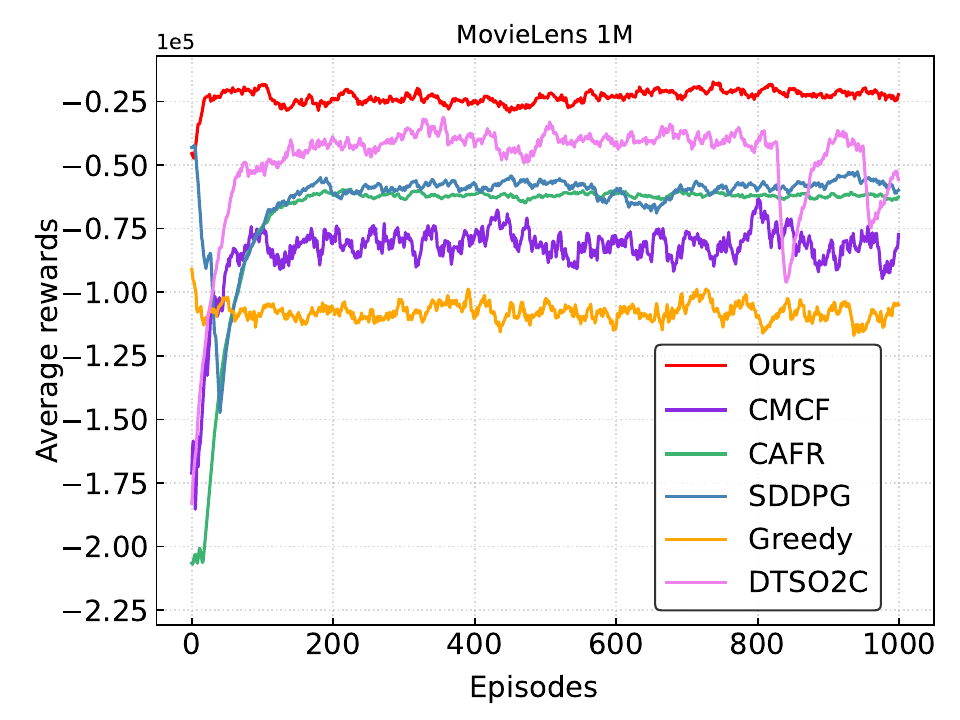}
        \footnotesize{(a) MovieLens 1M}
        \label{fig_7a}
    \end{minipage}
    \begin{minipage}{0.32\linewidth}
        \centering
        \includegraphics[width=\linewidth]{Fig/rewards/plottrain_rewards_curve.pdf}
        \footnotesize{(b) Top 5k}
        \label{fig_7b}
    \end{minipage}
    \begin{minipage}{0.32\linewidth}
        \centering
        \includegraphics[width=\linewidth]{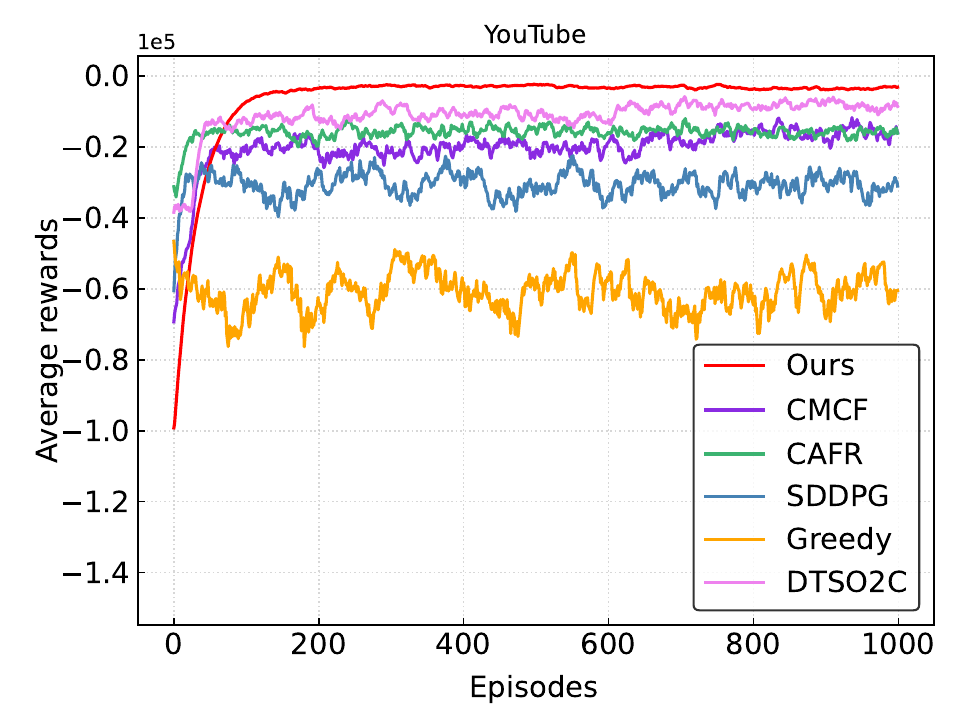}
        \footnotesize{(c) YouTube}
        \label{fig_7c}
    \end{minipage}
    \caption{Algorithm performance across different datasets.}
    \label{fig_algo}
\end{figure*}


Comparative experiments across three heterogeneous datasets (Table~\ref{def_dataset}) demonstrate that the proposed DAPR scheme significantly outperforms existing benchmark algorithms across three core metrics: transmission latency, cache hit rate, and cumulative reward. Specifically, on the MovieLens 1M dataset, DAPR achieved an average transmission latency of 30.86 ms, representing a 2.2\% reduction compared to the optimal baseline DTSO2C. The cache hit rate increased to 36.90\%, an 8.7\% improvement over the $\epsilon$-greedy baseline, while the aggregate reward value reached -22.51, marking a 4.5\% enhancement over the second-best CAFR algorithm. Performance gains are even more pronounced on the Top 5k dataset: transmission latency drops to 23.86 ms (2.7\% lower than DTSO2C), cache hit rate climbs to 56.84\% (an 8.7\% relative improvement over $\epsilon$-greedy), and the reward value of -15.47 represents a 3.4\% optimisation over the synchronous federated learning baseline DTSO2C. Test results on the YouTube dataset further validate the scheme's robustness, setting new benchmarks for latency, hit rate, and reward value. Specifically, the hit rate surpasses SDDPG by 1.2\%, while latency is reduced by 1.4\% compared to DTSO2C. Notably, the stable superior performance across datasets stems primarily from the GRU-VAE module's effective modelling of sparse request patterns, the asynchronous federated learning mechanism's dynamic adaptation to high vehicle mobility, and the SAC algorithm's precise trade-off in cache-latency joint optimisation. This fully demonstrates DAPR's technical feasibility and generalisability in real-world vehicle edge computing scenarios.

\begin{table}[htbp]
\begin{center}
\caption{Performance with Different Datasets}
\resizebox{\columnwidth}{!}{ 
\begin{tabular}{c c c c c c} 
\toprule
{Dataset} & {Algorithm} & { Delay (ms)} & {Hit Ratio} & { Reward } \\
\midrule
           & CMCF      & 32.44   & 35.01   & -24.23              \\
           & CAFR      & 31.94   & 34.90   & -23.57                 \\
MovieLens 1M  & SDDPG  & 32.15  & 34.51   & -24.32               \\
         & $\epsilon$-greedy   & 32.66   & 33.83  & -28.35               \\
         & DTSO2C     & 31.57  & 32.43   & -23.36                \\
           & Ours      & \textbf{30.86}  & \textbf{36.90}   & \textbf{-22.51}          \\
\midrule
           & CMCF       & 24.56    & 54.85   & -16.53              \\
           & CAFR     & 24.87    & 55.43   & -16.89                 \\
Top 5k       & SDDPG     & 24.96    & 54.42   & -16.25               \\
         & $\epsilon$-greedy     & 25.65    & 52.31   & -17.23       \\
         & DTSO2C     & 24.52  & 54.64   & -16.02                \\
           & Ours       & \textbf{23.86}    & \textbf{56.84}   & \textbf{-15.47}          \\
\midrule
           & CMCF       & 28.53    & 48.53   & -22.46              \\
           & CAFR     & 28.57    & 48.65   & -22.54                 \\
YouTube    & SDDPG     & 28.63    & 49.01   & -22.89               \\
        & $\epsilon$-greedy     & 29.13    & 47.63   & -24.53         \\
        & DTSO2C     & 27.86  & 48.98   & -22.32                \\
           & Ours       & \textbf{27.48}    & \textbf{49.58}   & \textbf{-20.57}          \\

\bottomrule
\end{tabular}
}
\label{def_dataset}
\end{center}
\end{table}

Fig.~\ref{fig_algo} illustrates the training convergence curves of various algorithms across different datasets, further revealing the inherent advantages of the proposed scheme in terms of dynamic stability and efficiency. Across the MovieLens 1M, Top 5k, and YouTube scenarios, DAPR's average reward value exhibits a rapid upward trajectory, converging to an optimal steady state within approximately 400 iterations—significantly earlier than the comparison algorithms. Specifically, DAPR achieves a final reward of -0.58 on the Top 5k dataset, surpassing CAFR and SDDPG by 19.4\% and 28.4\% respectively, while exhibiting lower variance, indicating superior robustness in policy learning. This performance advantage stems from the GRU-VAE architecture capturing latent space distributions through variational inference, effectively mitigating the sparsity and noise interference inherent in vehicle request data. Conversely, traditional autoencoders (employed in CMCF and CAFR) suffer from limited predictive accuracy due to their lack of temporal modelling capabilities, resulting in delayed caching decisions. Furthermore, the $\epsilon$-greedy strategy, failing to leverage historical request patterns, exhibits volatile convergence curves and ultimately mediocre performance, underscoring the necessity of reinforcement learning over heuristic methods when addressing complex environments. Outstanding cross-dataset consistency confirms that DAPR achieves adaptive optimisation for highly dynamic in-vehicle edge environments through real-time state awareness provided by the digital twin layer and dynamic weight aggregation via asynchronous federated learning.

\begin{figure}
    \centering
    \includegraphics[width=0.85\linewidth]{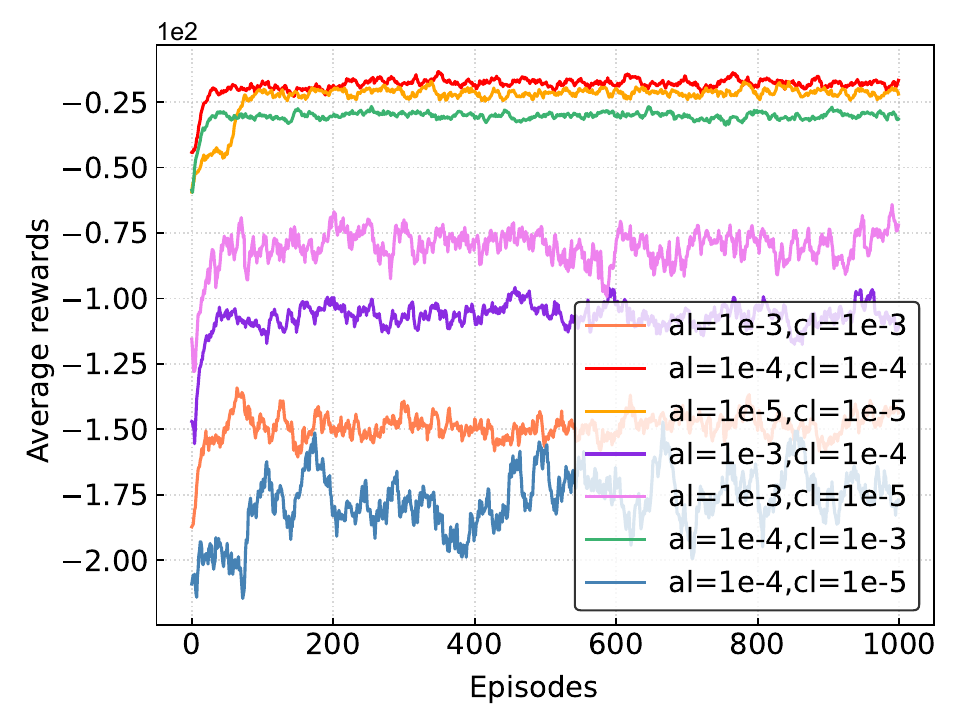}
    \caption{Performance of SAC algorithm training at different learning rates.}
    \label{fig_learn}
\end{figure}

In complex scenarios, a reasonable learning rate setting can significantly improve the stability and convergence speed of the algorithm. We tested the effect of the SAC algorithm on the most average reward under different actor network learning rates and different critic network learning rates. As shown in Fig.~\ref{fig_learn}, the algorithm converges to the highest reward value at $al=10^{-4}$ and $cl=10^{-4}$. In contrast, the remaining cases, while improving initially, end up with lower stable rewards than the $al=10^{-4}$ and $cl=10^{-4}$ cases and show slower convergence and poorer final performance. Therefore, when setting the learning rate, we choose $al=10^{-4}$ and $cl=10^{-4}$.


\begin{figure*}[!htbp]
    \centering
    \begin{minipage}{0.32\linewidth}
        \centering
        \includegraphics[width=\linewidth]{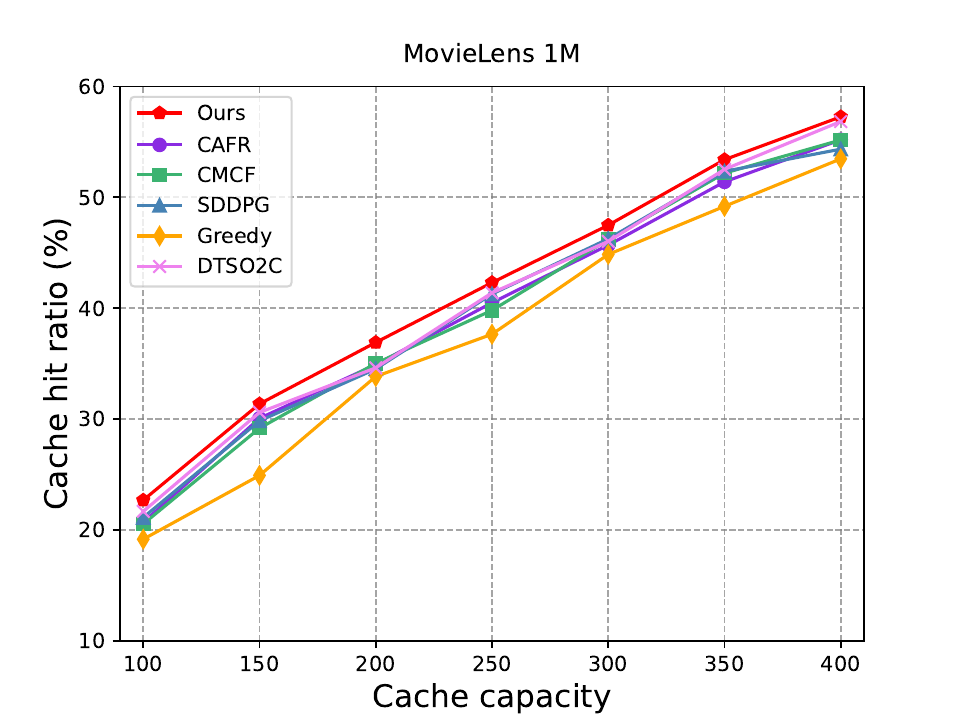}
        \footnotesize{(a) MovieLens 1M}
        \label{fig_9a}
    \end{minipage}
    \begin{minipage}{0.32\linewidth}
        \centering
        \includegraphics[width=\linewidth]
        {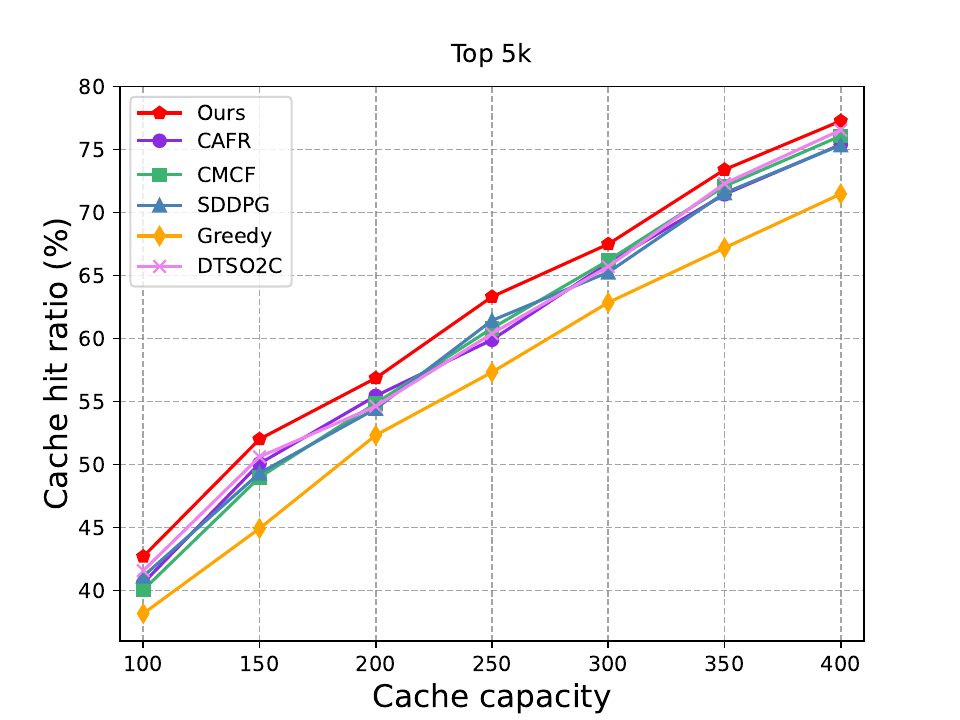}
        \footnotesize{(b) Top 5k}
        \label{fig_9b}
    \end{minipage}
    \begin{minipage}{0.32\linewidth}
        \centering
        \includegraphics[width=\linewidth]
        {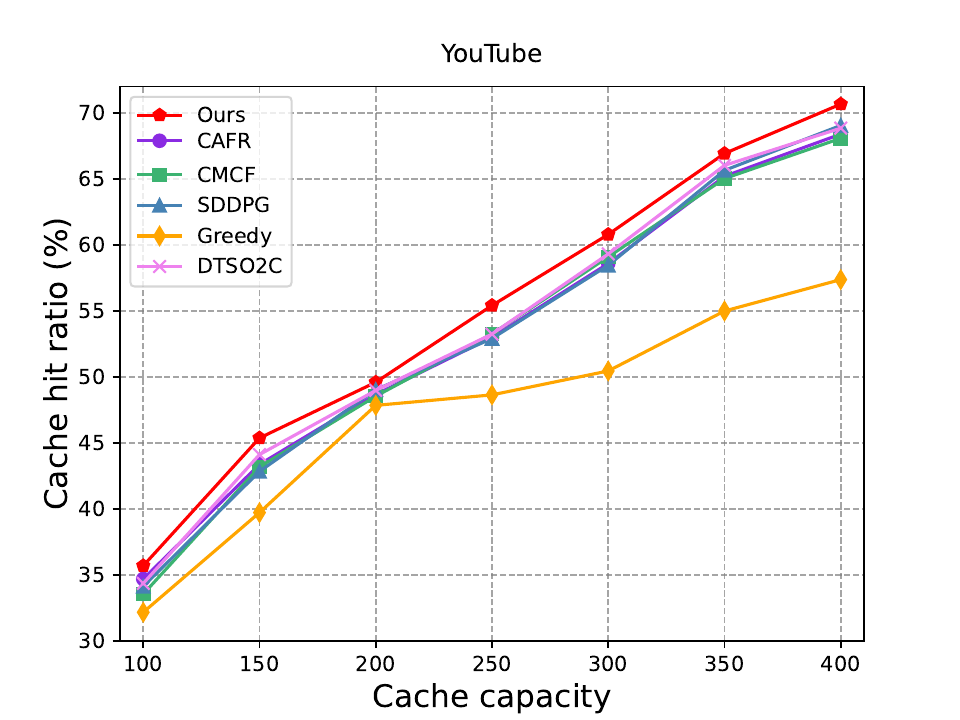}
        \footnotesize{(c) YouTube}
        \label{fig_9c}
    \end{minipage}
    \caption{Impact of cache capacity and algorithm effects on cache hit rate across different datasets.}
    \label{fig_algo_eff}
\end{figure*}


\begin{figure*}[!htbp]
    \centering
    \begin{minipage}{0.32\linewidth}
        \centering
        \includegraphics[width=\linewidth]
        {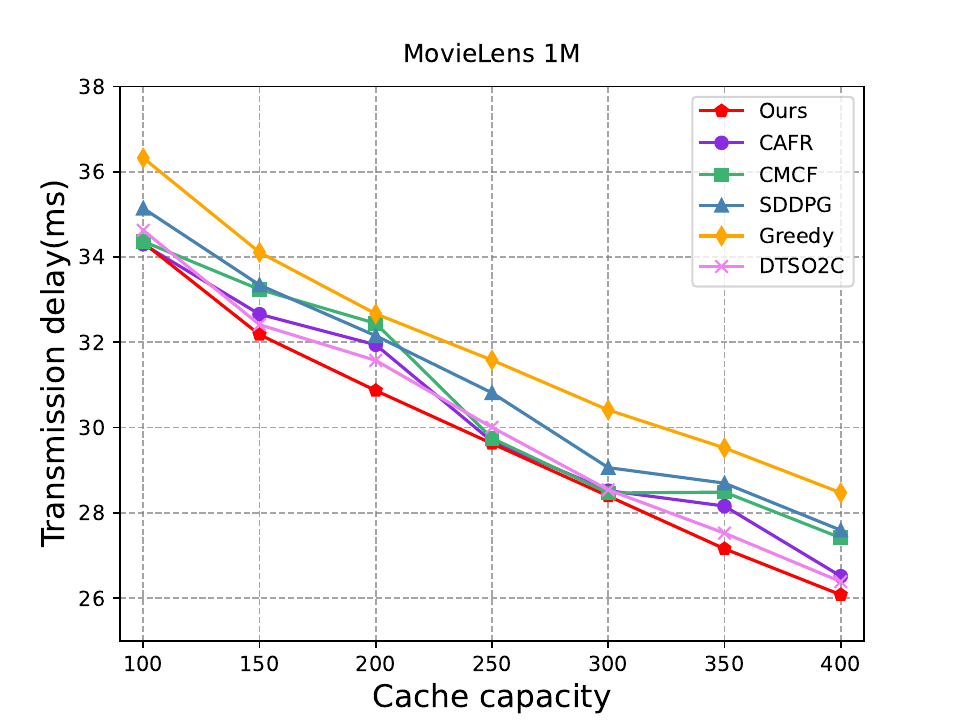}
        \footnotesize{(a) MovieLens 1M}
        \label{fig_10a}
    \end{minipage}
    \begin{minipage}{0.32\linewidth}
        \centering
        \includegraphics[width=\linewidth]
        {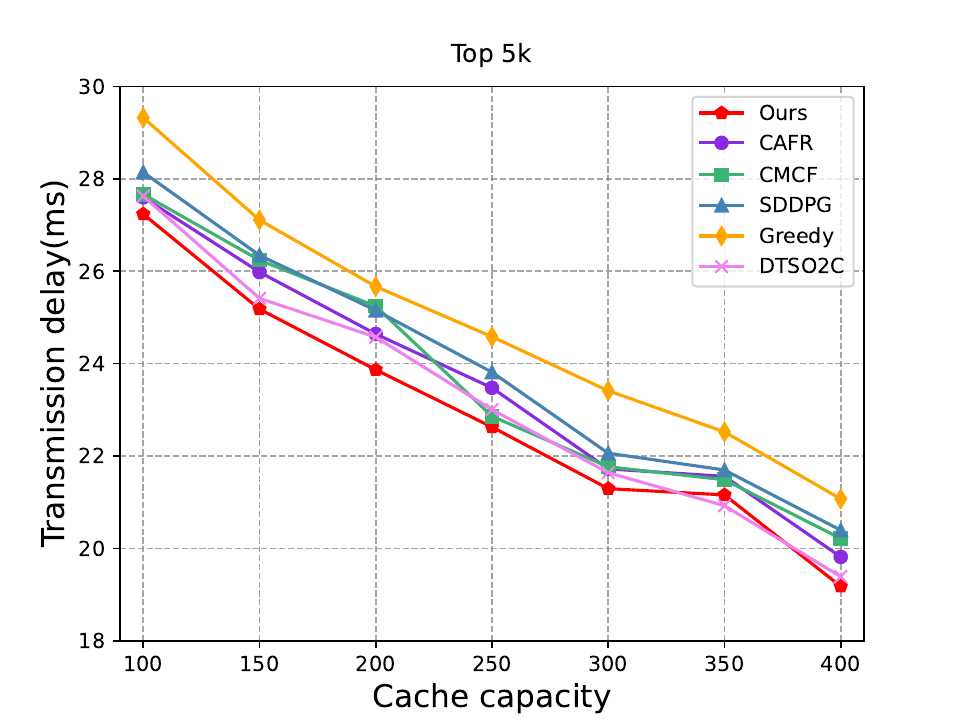}
        \footnotesize{(b) Top 5k}
        \label{fig_10b}
    \end{minipage}
    \begin{minipage}{0.32\linewidth}
        \centering
        \includegraphics[width=\linewidth]
        {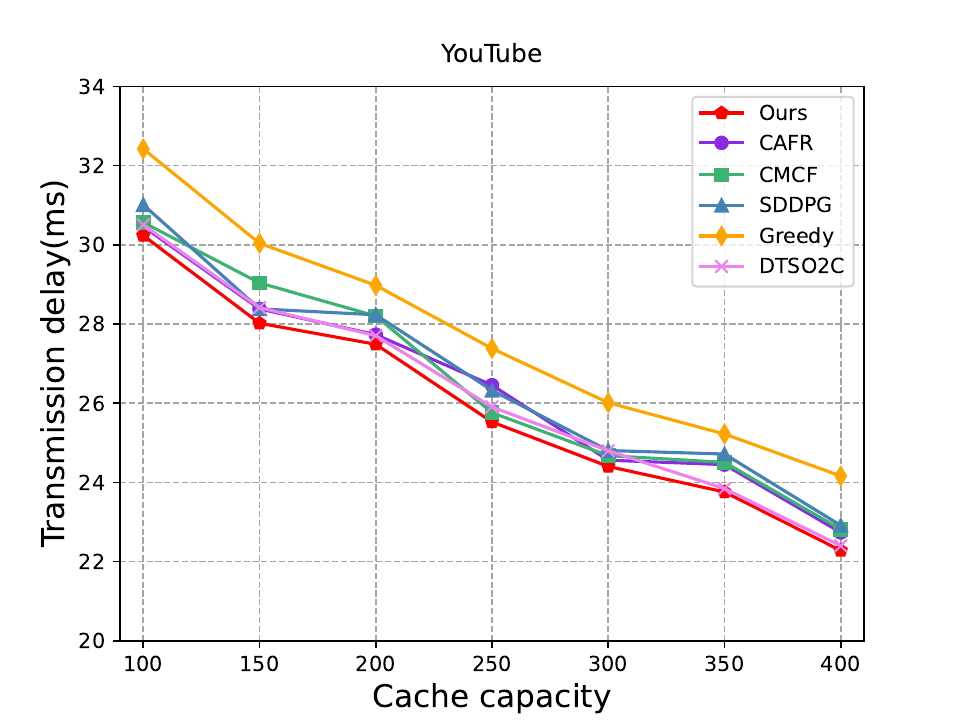}
        \footnotesize{(c) YouTube}
        \label{fig_10c}
    \end{minipage}
    \caption{Cache hit ratio for each algorithm with different cache capacities.}
    \label{fig_algo_ts}
\end{figure*}


Fig.~\ref{fig_algo_eff} and~\ref{fig_algo_ts} detail the impact of varying cache capacities on the cache hit rate and transmission latency of each algorithm across three datasets: MovieLens 1M, Top 5k, and YouTube. Experimental results demonstrate that as cache capacity increases, the cache hit rate improves for all algorithms, whilst transmission latency correspondingly decreases. This trend aligns with expectations, as larger cache capacities store more content, thereby increasing the probability of local cache hits, reducing requests for remote content, and consequently lowering transmission latency. Nevertheless, significant performance differences persist between the algorithms. The DAPR algorithm demonstrated superior performance across varying cache capacities. As cache size increased, DAPR exhibited markedly higher cache hit rates alongside correspondingly reduced transmission latency. On the MovieLens 1M dataset, DAPR achieved a cache hit rate of about 60\% at high cache capacities, with transmission latency dropping to approximately 26ms. On the Top 5k dataset, DAPR achieved a cache hit rate nearing 75\% at high capacities, with transmission latency reduced to approximately 17ms. On the YouTube dataset, DAPR surpassed a 70\% cache hit rate, reducing transmission latency to approximately 22ms. These results validate DAPR's effectiveness in enhancing cache performance, demonstrating its robustness and superiority across varying cache capacities.

\begin{figure*}[!htbp]
    \centering
    \begin{minipage}{0.32\linewidth}
        \centering
        \includegraphics[width=\linewidth]{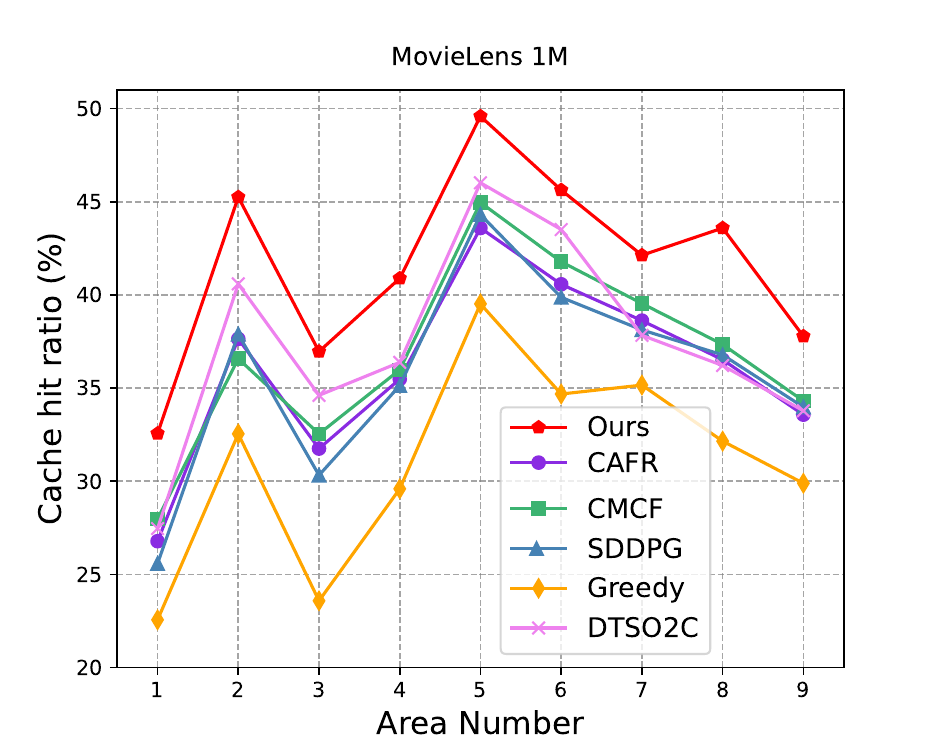}
        \footnotesize{(a) MovieLens 1M}
        \label{fig_11a}
    \end{minipage}
    \begin{minipage}{0.32\linewidth}
        \centering
        \includegraphics[width=\linewidth]
        {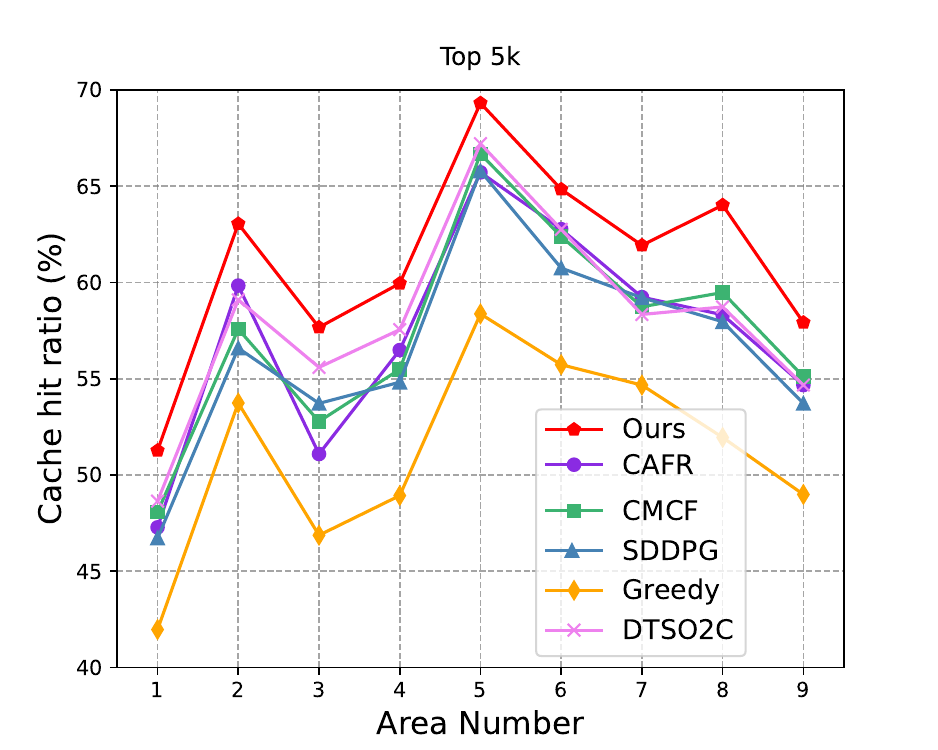}
        \footnotesize{(b) Top 5k}
        \label{fig_11b}
    \end{minipage}
    \begin{minipage}{0.32\linewidth}
        \centering
        \includegraphics[width=\linewidth]
        {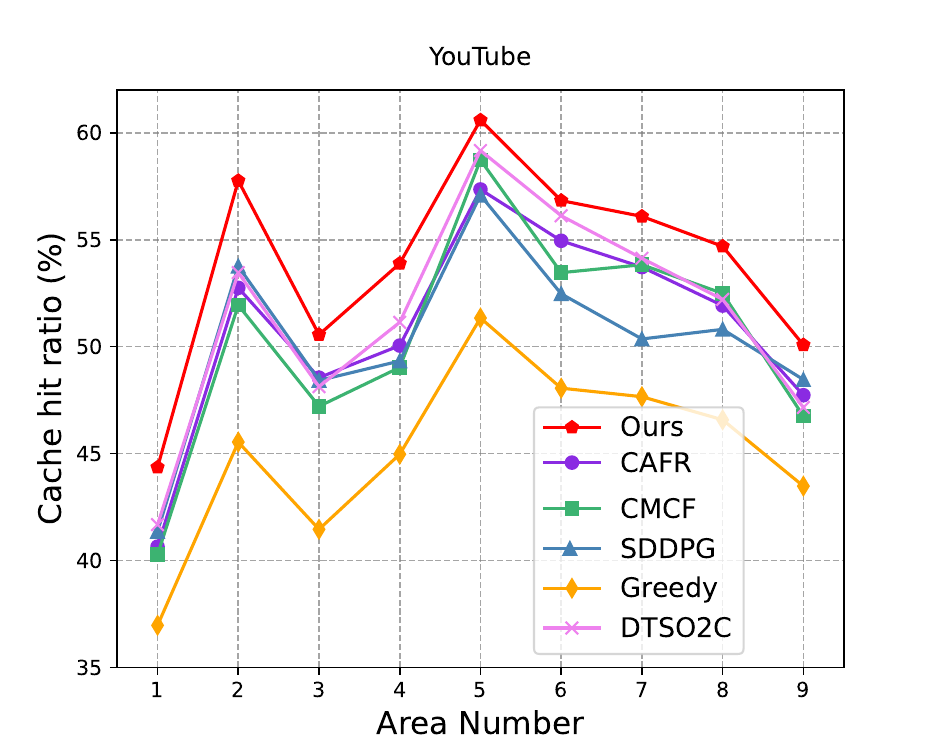}
        \footnotesize{(c) YouTube}
        \label{fig_11c}
    \end{minipage}
    \caption{Average cache hit ratio of each area under different schemes.}
    \label{fig_eff_area}
\end{figure*}



\begin{figure*}[!htbp]
    \centering
    \begin{minipage}{0.32\linewidth}
        \centering
        \includegraphics[width=\linewidth]{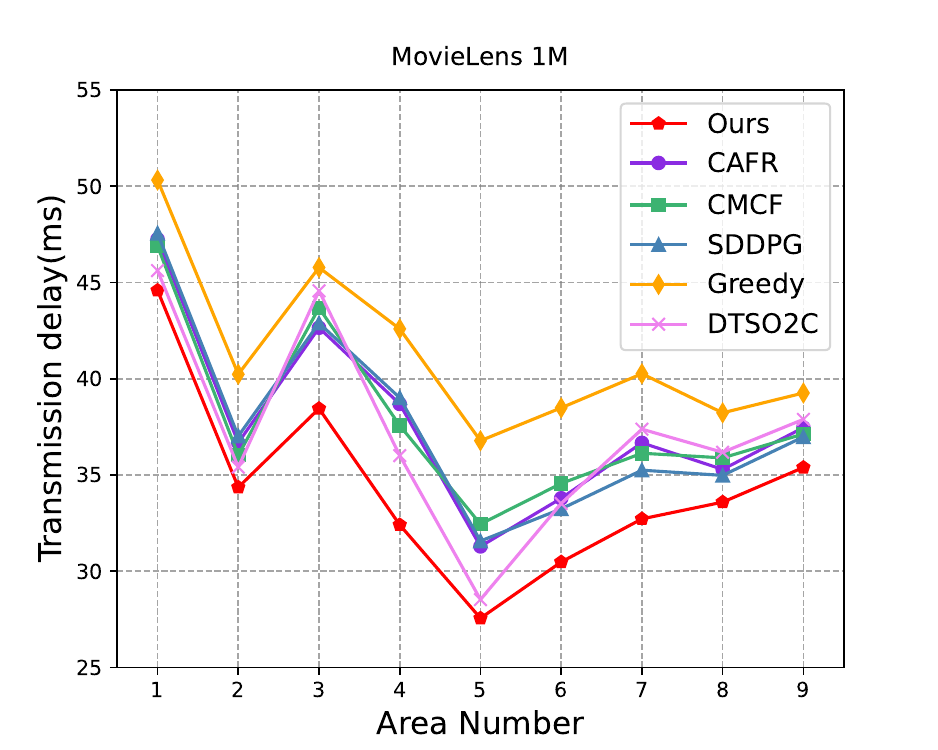}
        \footnotesize{(a) MovieLens 1M}
        \label{fig_12a}
    \end{minipage}
    \begin{minipage}{0.32\linewidth}
        \centering
        \includegraphics[width=\linewidth]
        {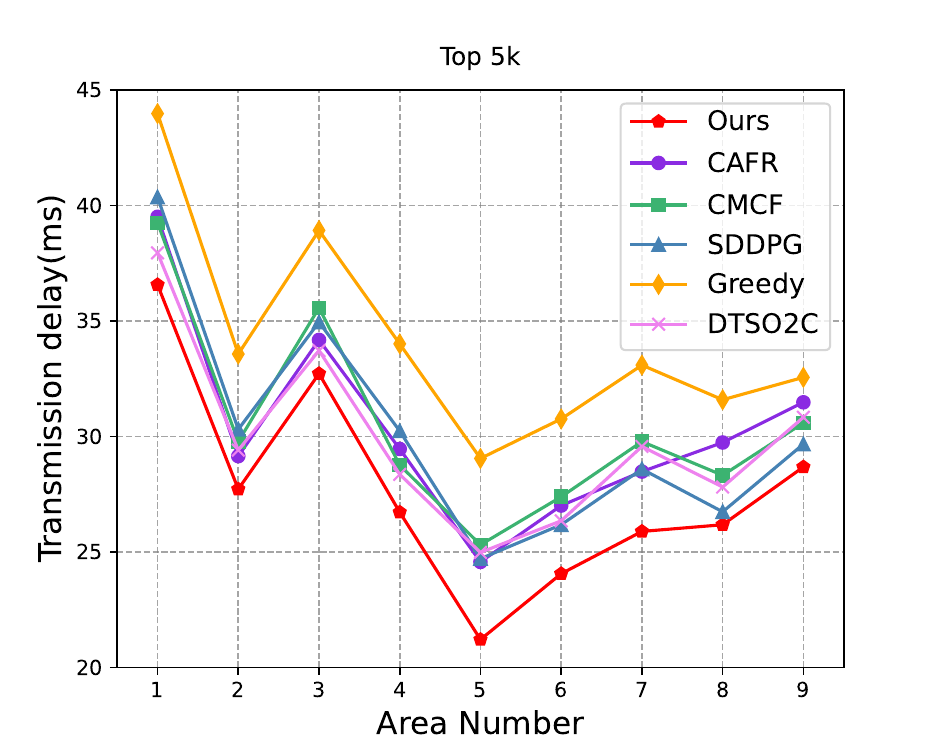}
        \footnotesize{(b) Top 5k}
        \label{fig_12b}
    \end{minipage}
    \begin{minipage}{0.32\linewidth}
        \centering
        \includegraphics[width=\linewidth]
        {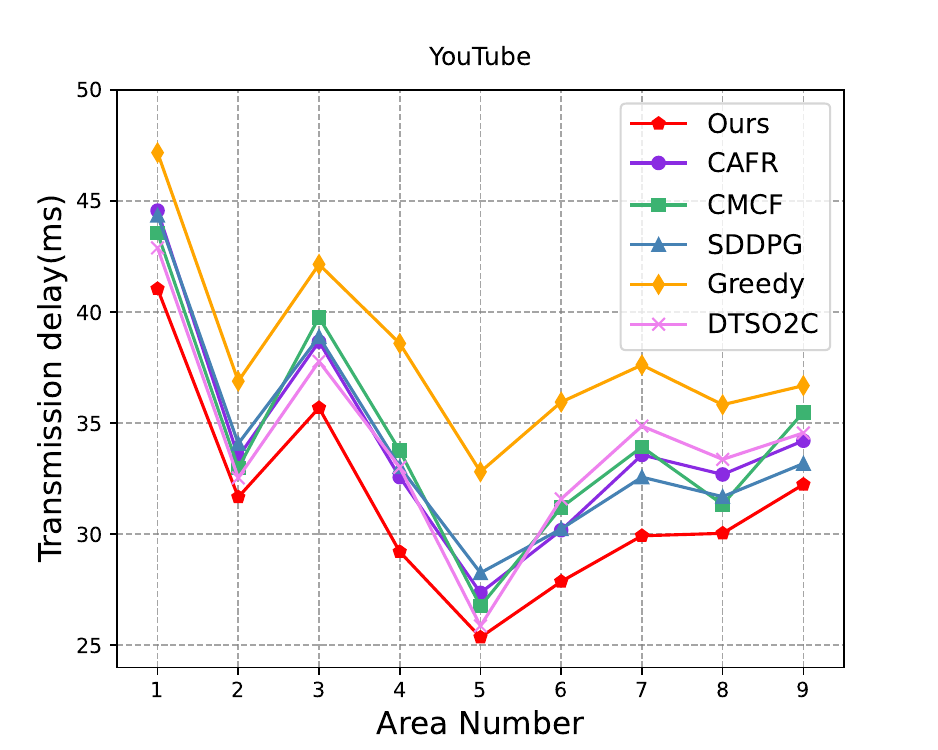}
        \footnotesize{(c) YouTube}
        \label{fig_12c}
    \end{minipage}
    \caption{Average delay of each area under different schemes.}
    \label{fig_delay_area}
\end{figure*}

Fig.~\ref{fig_eff_area} and~\ref{fig_delay_area} illustrate the average cache hit rate and average transmission latency across different regions for various caching strategies under three datasets (MovieLens 1M, Top 5k, YouTube), validating the impact of regional user density heterogeneity on strategy performance. In low-density zones (1–3), all strategies exhibit low hit rates (below 30\% for MovieLens 1M, below 35\% for Top 5k and YouTube) due to insufficient cache reuse caused by sparse user distribution. Concurrently, transmission latency remains high as RSUs lack sufficient training data for precise forecasting, necessitating frequent remote content retrieval. Performance significantly improves in medium-to-high density zones (4–9): Our approach achieves a hit rate exceeding 56\% in Top 5k region 4 and maintains approximately 25 milliseconds latency in YouTube regions 8–9, benefiting from enhanced cache synergies through abundant user requests. Across all scenarios, our strategy outperforms alternatives—reaching a peak hit rate of 58\% in Top 5k region 6, with latency 3–5 milliseconds lower than CAFR/CMCF in high-density areas and minimal dataset variability. Benchmark strategies (CAFR, CMCF) perform well in medium-to-high density zones but degrade in low-density areas. The greedy strategy fared worst, with sub-25\% hit rates in low-density zones and YouTube latency fluctuations exceeding 15 milliseconds. Cross-dataset stability confirms superior performance and exceptional density adaptability when handling heterogeneous user density distributions. DAPR adapts to diverse content patterns, providing practical support for in-vehicle edge computing zone caching deployment.

\begin{figure*}[htbp]
    \centering
    \begin{minipage}{0.32\linewidth}
        \centering
        \includegraphics[width=\linewidth]{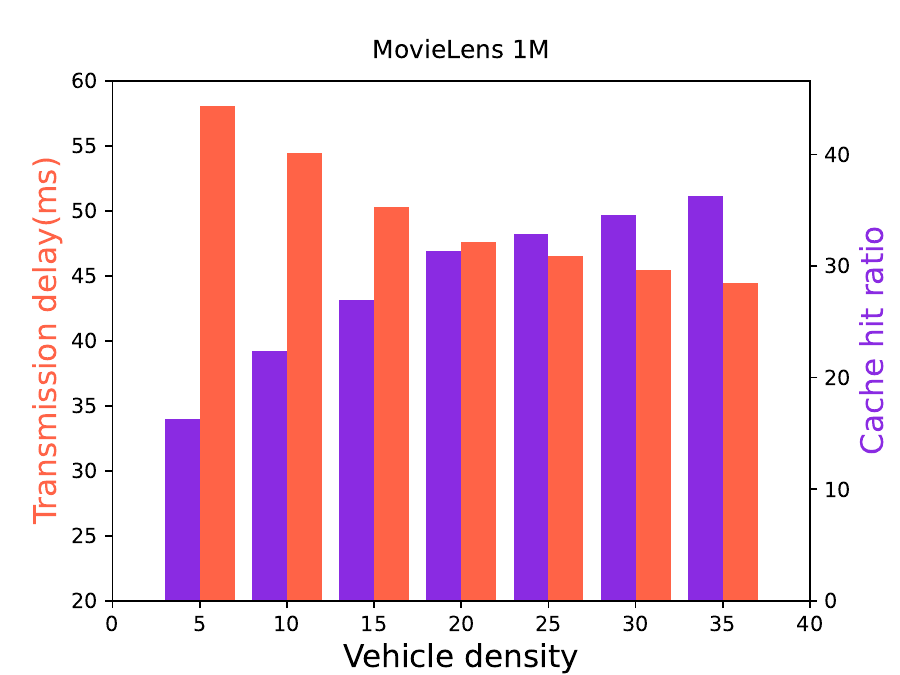}
        \footnotesize{(a) MovieLens 1M}
        \label{fig_13a}
    \end{minipage}
    \begin{minipage}{0.32\linewidth}
        \centering
        \includegraphics[width=\linewidth]
        {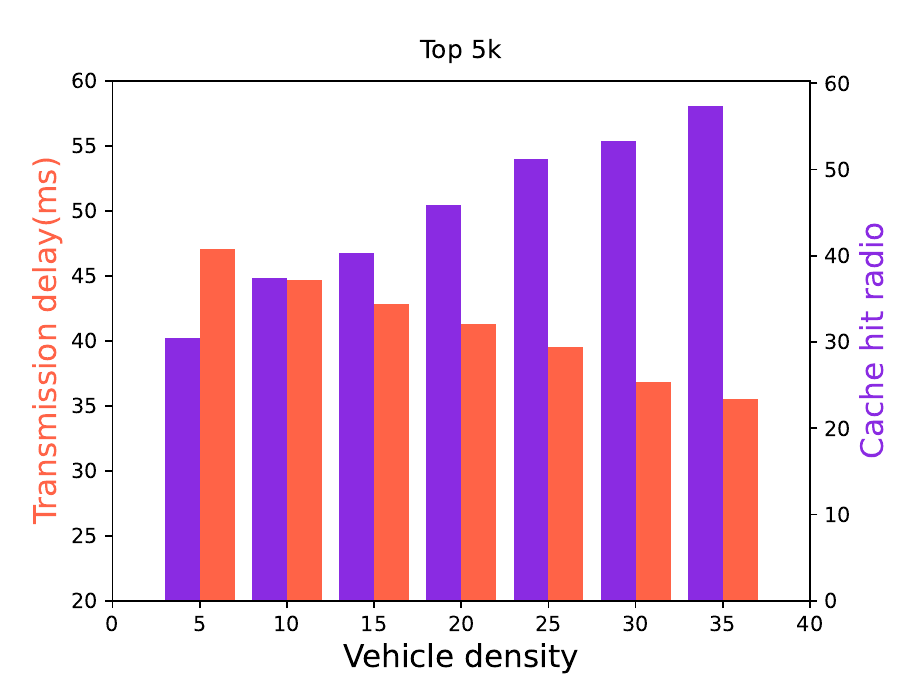}
        \footnotesize{(b) Top 5k}
        \label{fig_13b}
    \end{minipage}
    \begin{minipage}{0.32\linewidth}
        \centering
        \includegraphics[width=\linewidth]
        {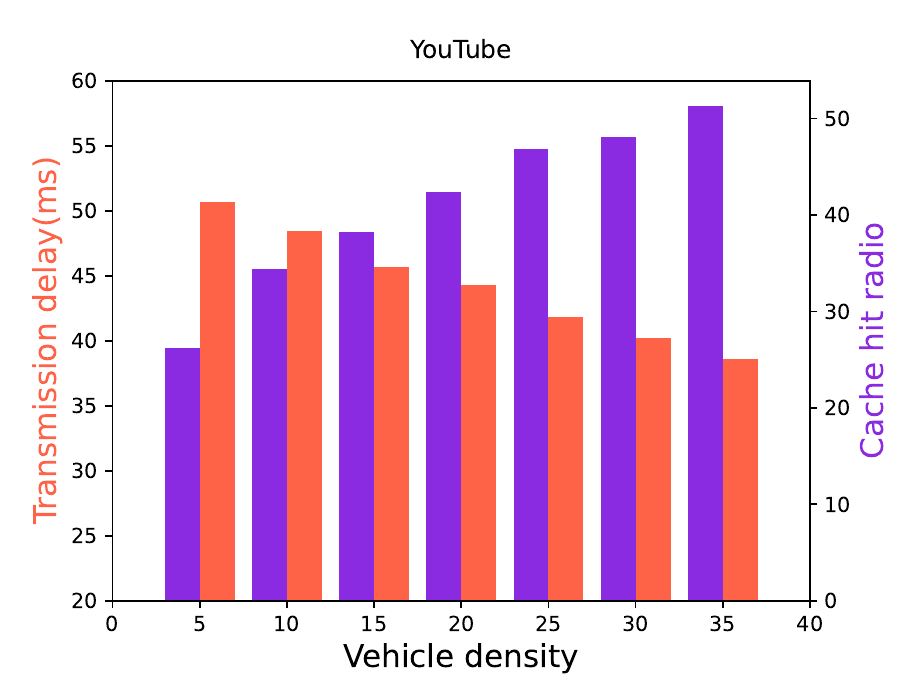}
        \footnotesize{(c) YouTube}
        \label{fig_13c}
    \end{minipage}
    \caption{Performance comparison at different vehicle densities.}
    \label{fig_debsity}
\end{figure*}

Fig.~\ref{fig_debsity} presents the performance test results for our approach across the MovieLens 1M, Top 5k, and YouTube datasets, demonstrating cache hit rates and content delivery latency under varying vehicle densities. Analysis indicates that as vehicle density increases, cache hit rates improve while content delivery latency progressively decreases. This occurs because higher vehicle density entails more frequent access to the current Roadside Unit (RSU), enabling the RSU to gather more training samples and capture additional potential popular content. Consequently, it can more accurately predict future popular content, allowing more vehicles to access content directly from the local RSU. All three datasets exhibit similar trends, confirming our solution maintains performance gains across diverse content types and user behaviour patterns. This cross-dataset consistency underscores the robustness and adaptability of our approach, demonstrating its capacity to dynamically adjust to environmental changes and leverage increased vehicle density to optimise caching strategies, thereby enhancing overall system performance.


\begin{figure*}[htbp]
    \centering
    \begin{minipage}{0.32\linewidth}
        \centering
        \includegraphics[width=\linewidth]{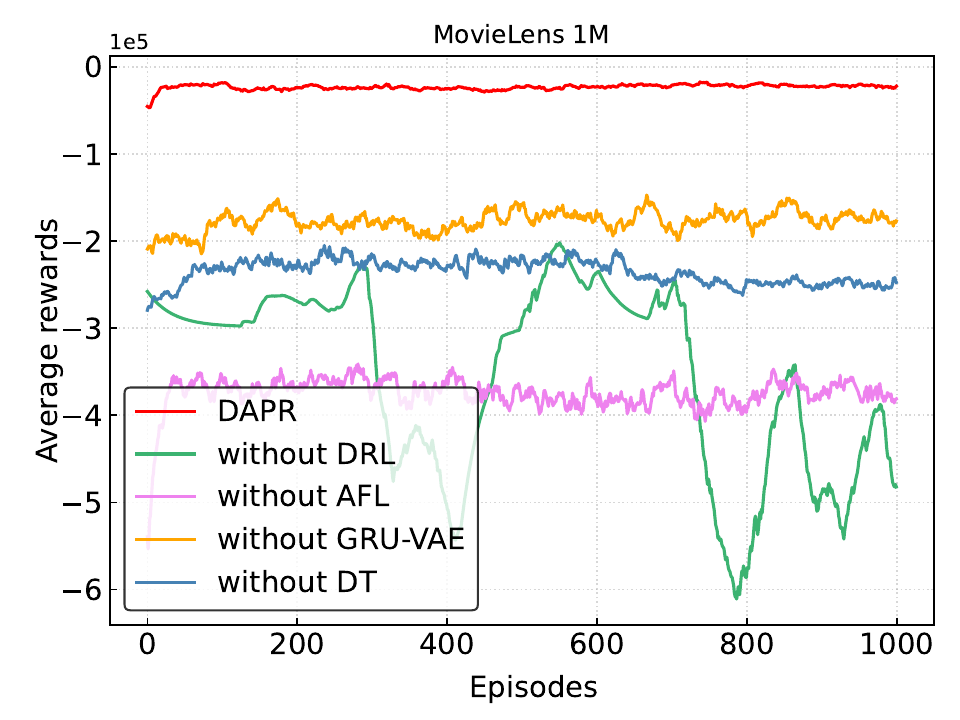}
        \footnotesize{(a) MovieLens 1M}
        \label{fig_15a}
    \end{minipage}
    \begin{minipage}{0.32\linewidth}
        \centering
        \includegraphics[width=\linewidth]
        {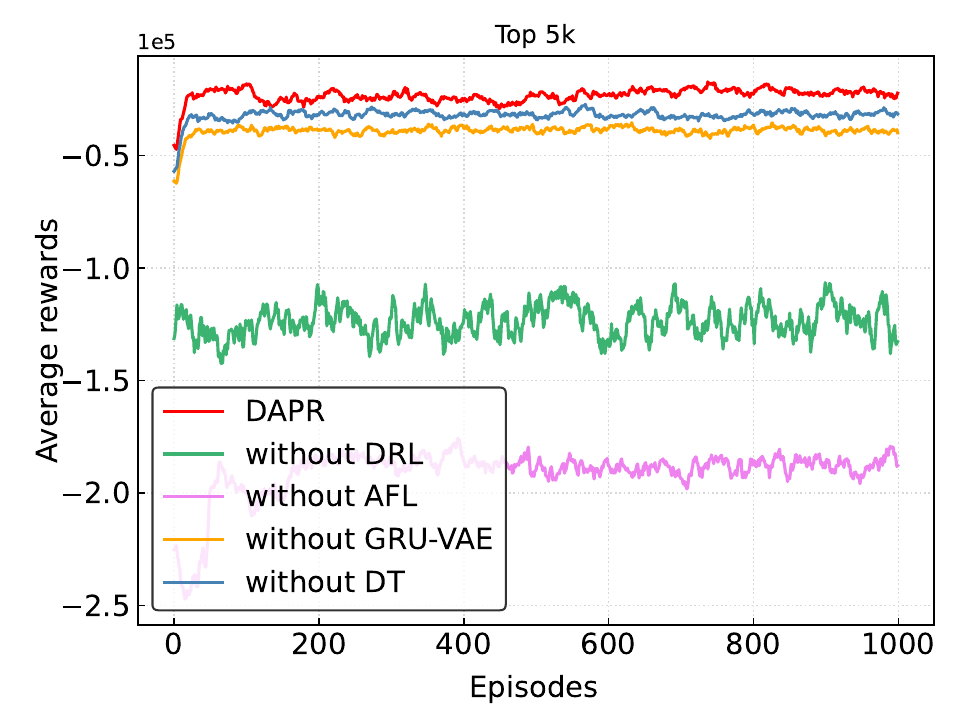}
        \footnotesize{(b) Top 5k}
        \label{fig_15b}
    \end{minipage}
    \begin{minipage}{0.32\linewidth}
        \centering
        \includegraphics[width=\linewidth]
        {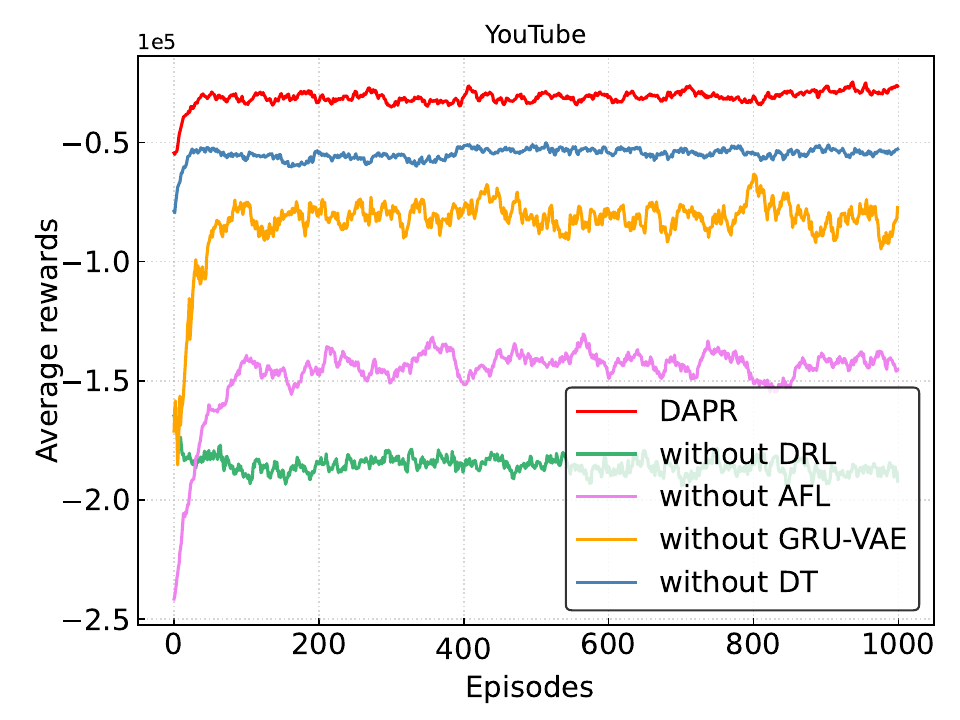}
        \footnotesize{(c) YouTube}
        \label{fig_15c}
    \end{minipage}
    \caption{Comparative experiments on critical module ablation in dynamic edge caching scenarios.}
    \label{fig_with}
\end{figure*}

\begin{figure}
    \centering
    \includegraphics[width=0.9\linewidth]{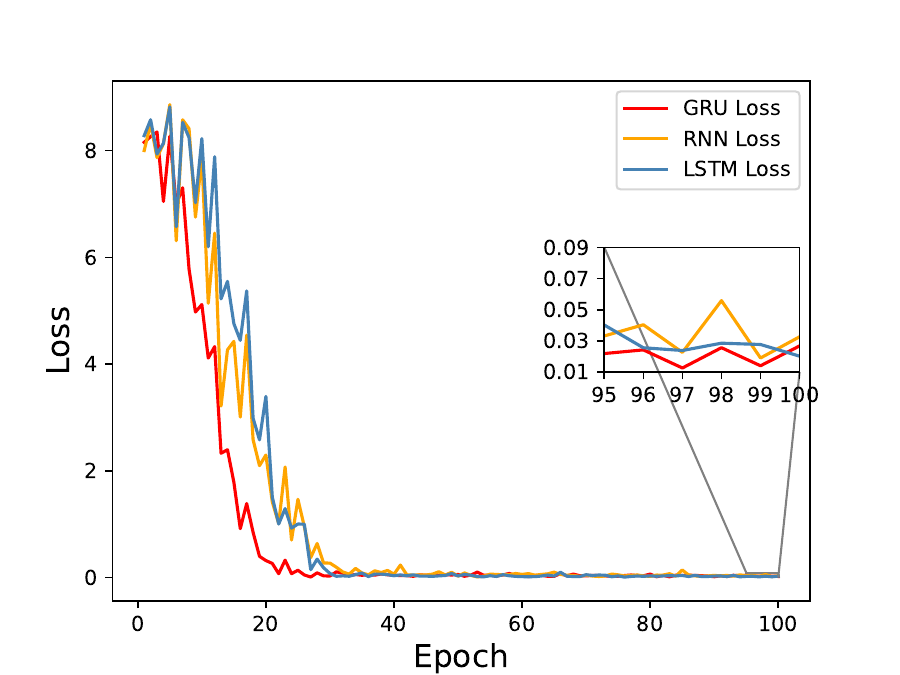}
    \caption{Comparison with LSTM and RNN predicted loss.}
    \label{fig_loss}
\end{figure}

To evaluate the prediction results of our content popularity prediction, we compare the MSE of GRU-based, RNN-based, and LSTM-based predictions, as seen in Fig.~\ref{fig_loss}, the loss value of GRU decreases faster in the early stage and converges to a lower level faster in the late stage of training. This is due to the time-series nature of the data and the complexity of the environment in vehicle scenarios, as well as the limitations of the computing resources and time of the devices. GRU can adapt to highly dynamic and resource-constrained scenarios more efficiently than RNN and LSTM due to its streamlined gating structure and fewer parameters than RNN and LSTM, which is suitable for the real-time inference needs of the edge devices, and it has a short and stable path of the gradient propagation to alleviate the problem of gradient disappearance in the prediction of the vehicle track under the long sequences. The short and stable gradient propagation path alleviates the gradient vanishing problem in long sequence vehicle trajectory prediction and realizes fast convergence.

To explore the role of each key component, we conducted ablation experiments with the following experimental setup:
\begin{itemize}
    \item without DRL: After obtaining popular content, the vehicle directly caches popular content randomly without selecting the optimal caching policy via DRL.
    \item  without AFL: Instead of using asynchronous federated learning, a federated averaging algorithm is used.
    \item without GRU-VAE: There is no content prediction function, and the system adopts the traditional way of determining the cached content based on historical access statistics.
    \item  without DT: In this configuration, the system does not utilise digital twin models to map changes in the physical network's state in real time. Instead, it relies solely on traditional methods for environmental perception and decision-making, without the support of digital twin technology.
\end{itemize}

The experimental results are shown in Fig.~\ref{fig_with}. Removing deep reinforcement learning leads to more volatility and lower rewards. In the dynamic edge caching environment for vehicles, conditions and demands are constantly changing. Random caching methods struggle to adapt, reducing cache hit rates and fulfilling user requests promptly. This results in a significant drop in average reward, highlighting the importance of reinforcement learning in caching decisions.

Removing asynchronous federated learning also decreases average rewards. It requires all nodes to complete training before uploading data, which can be delayed due to vehicle mobility. This delay prevents the model from reflecting the latest user demands and content trends, leading to mismatches between cached content and user needs and thus lower rewards.
Removing the GRU-VAE module results in a decline in average reward and a flattening curve. The vehicle environment has temporal characteristics in content requests, with varying user demands over time. Without the GRU-VAE module, the model cannot analyze temporal data deeply, relying only on surface information to predict content popularity. This reduces prediction accuracy, leading to a lower alignment between cached content and user demand, and ultimately a lower average reward.
Removing the digital twin also decreases average rewards, showing its role in providing real-time environmental awareness for cache deployment decisions. The digital twin model supports optimized asynchronous federated learning and cache deployment by mapping physical network state changes.
In summary, each component of the DAPR framework is essential for vehicle edge caching performance. The integration of deep reinforcement learning, asynchronous federated learning, the GRU-VAE prediction model, and digital twin technology allows the framework to make optimal caching decisions in dynamic environments, maximizing benefits and minimizing latency.

\subsection{Discussion}
The DAPR framework exhibits strong versatility in VEC scenarios. At the data level, it adapts to heterogeneous datasets and diverse multimedia request patterns, yielding superior performance in terms of cache hit rates, latency, and cumulative rewards. At the environmental level, it robustly handles variations in vehicle distribution, spanning from low-density suburbs to high-density urban centers. At the resource level, its architecture is compatible with the constrained resources of RSUs and in-vehicle units in distributed training modes, making it highly suitable for real-world deployment.

Nevertheless, this framework still exhibits certain limitations in practical application. Firstly, multi-module joint optimization demands high coordination in parameter updates, which can impede convergence speed under low-density sparse data scenarios. Secondly, the framework currently lacks protective measures against security threats such as malicious node attacks and data privacy breaches. To address these challenges, subsequent work will focus on optimization across two dimensions: First, leveraging federated distillation to achieve effective knowledge transfer from global to local models, supplemented by generative data augmentation to expand sample sizes. This dual-pronged approach of knowledge sharing and data augmentation aims to overcome the constraints on convergence speed imposed by sparse data. Second, incorporating blockchain technology to establish a decentralized, trusted parameter aggregation mechanism, alongside employing differential privacy techniques to enhance data security within heterogeneous networks. This will foster a more efficient and robust federated optimization framework.

\section{Conclusion}\label{Se:6}

In this paper, we focus on the edge caching optimization problem in digital twin-enabled vehicular edge computing scenarios, and propose a DAPR scheme to address the caching challenges caused by vehicle mobility and network dynamics changes, as well as user data privacy protection issues. The scheme integrates asynchronous federated learning, spatio-temporal feature generation and deep reinforcement learning techniques to construct a collaborative framework for privacy protection and performance optimization. By designing an asynchronous federated learning mechanism based on dynamic weight allocation, the communication overhead is reduced while protecting user data privacy; the accuracy of dynamic content popularity prediction is improved by using the GRU-VAE dual-channel feature extraction mechanism; and the online caching decision-making model constructed based on the SAC algorithm maximizes the long-term caching revenue, significantly improves the cache hit ratio, and reduces the transmission latency. In addition, the superiority of the scheme is also demonstrated experimentally.

\bibliographystyle{IEEEtran}
\bibliography{ref}

\begin{IEEEbiography}
[{\includegraphics[width=1in,height=1.25in,clip,keepaspectratio]
{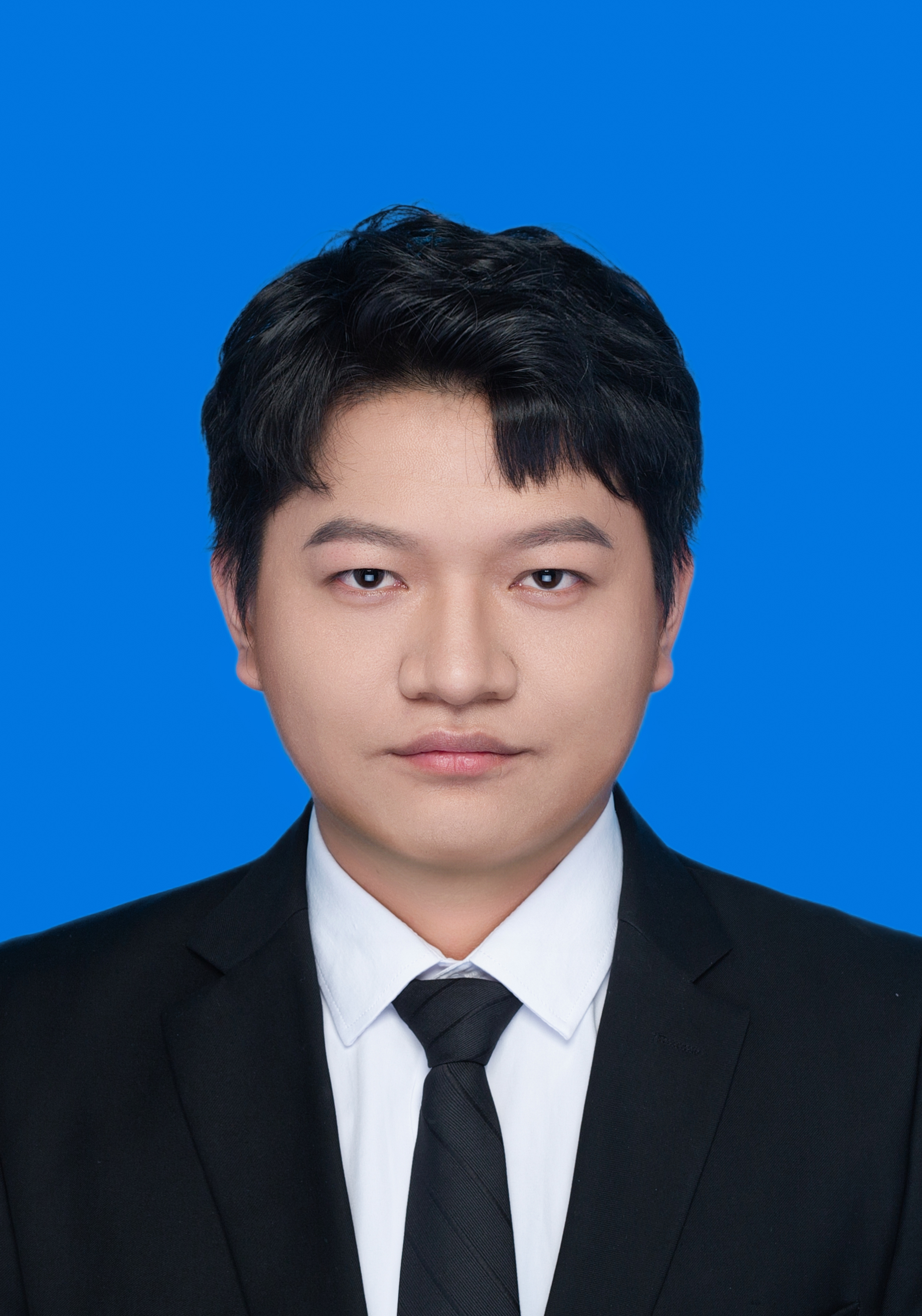}}]
{Jiahao Zeng}
is currently pursuing the Master's degree with the School of Computer Science and Engineering, Guangxi Normal University, Guilin, China.

His research interests include edge computing, reinforcement learning, Internet of Things, and blockchain system.
\end{IEEEbiography}
\vspace{-1cm}
\begin{IEEEbiography}[{\includegraphics[width=1in,height=1.25in,clip,keepaspectratio]{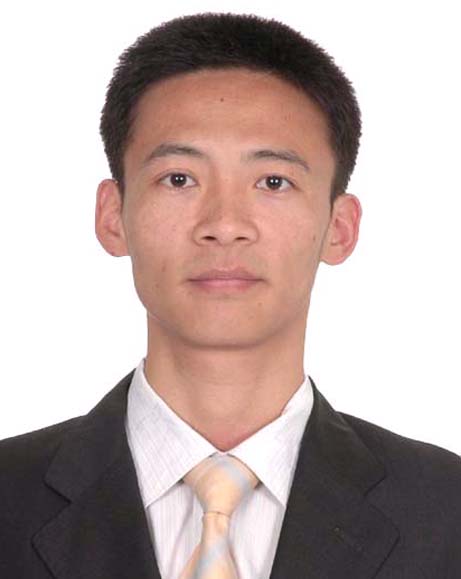}}]
{Zhenkui Shi}
is an associate professor at the Faculty of School of Computer Science and Engineering and School of Software, Guangxi Normal University, China. He received the Ph.D. degree in Computer Science from City University of Hong Kong in 2018. His research interests include data security and privacy, blockchain technology, IoT privacy and security, and Trustworthy AI.
\end{IEEEbiography}
\vspace{-1cm}
\begin{IEEEbiography}[{\includegraphics[width=1in,height=1.25in,clip,keepaspectratio]{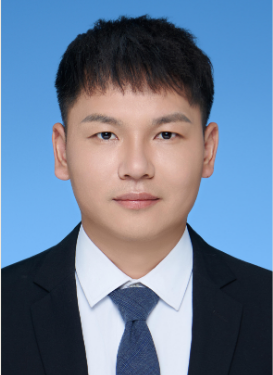}}]
    {Chunpei Li} received his Ph.D. from the School of Computer Science and Engineering at Guangxi Normal University in 2024. He is currently conducting postdoctoral research at the Ministry of Education Key Laboratory of Educational Blockchain and Intelligent Technology at Guangxi Normal University. His research interests include blockchain, artificial intelligence, and information security.
\end{IEEEbiography}
\vspace{-1cm}
\begin{IEEEbiography}[{\includegraphics[width=1in,height=1.25in,clip,keepaspectratio]{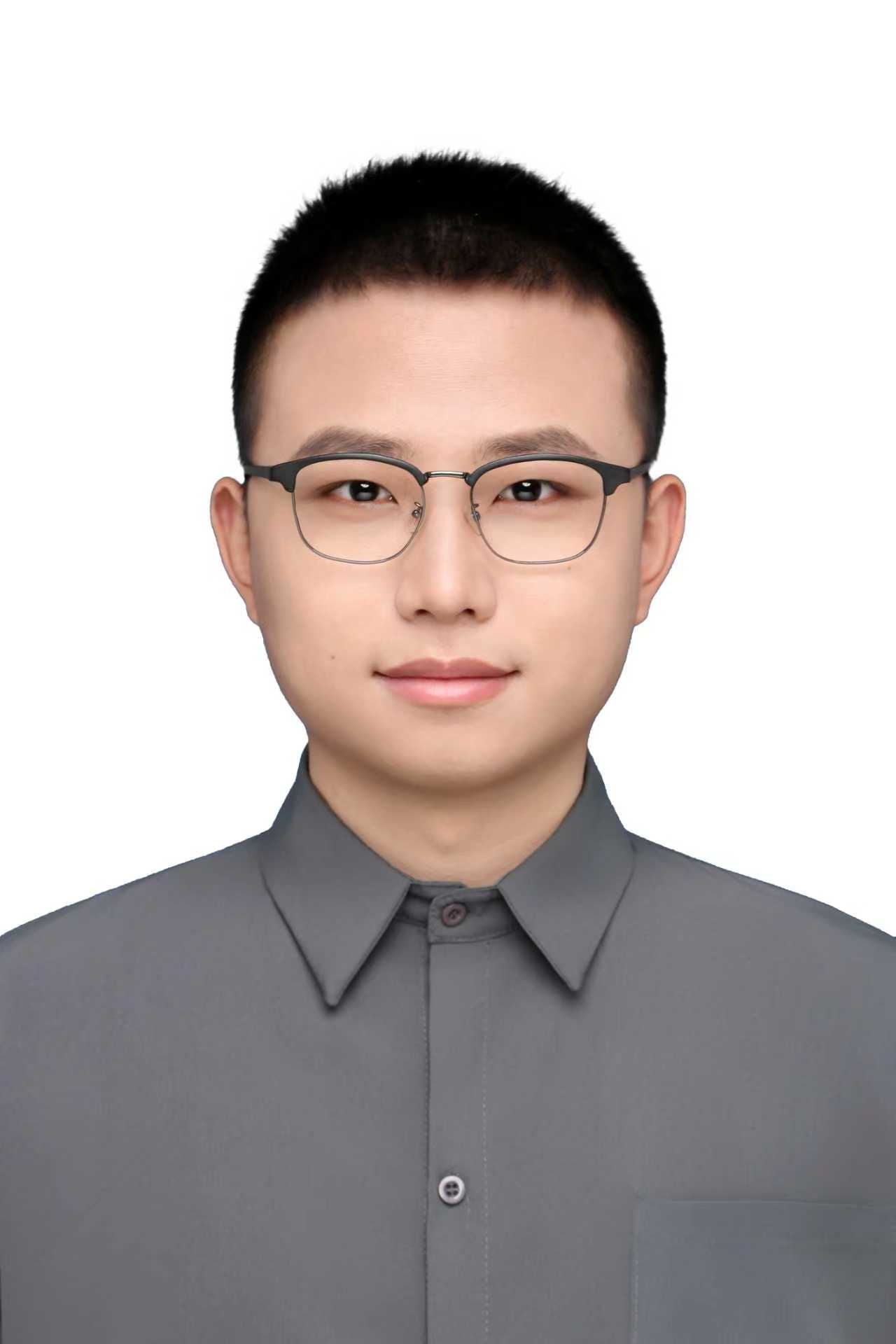}}]
{Mengkai Yan} received his Ph.D. in Control Science and Engineering from Nanjing University of Science and Technology in 2025, with a research focus on computer vision. He is currently a lecturer at Hohai University. His current research interests include computer vision and anomaly detection.

\end{IEEEbiography}
\vspace{-1cm}
\begin{IEEEbiography}[{\includegraphics[width=1in,height=1.25in,clip,keepaspectratio]{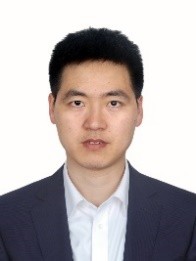}}]
{Hongliang Zhang} received the B.S. degree in Computer Science and Technology from Xiangtan University in 2010, and the M.S. degree in Computer Science and Technology from Xiangtan University in 2013. He is currently pursuing the Ph.D. degree in Computer Science and Technology from the Nanjing University of Science and Technology. His research interests include optimal transport, evolutionary computation, neural combinatorial optimization, multi-objective optimization, and medical image analysis. 

He has served as an PC Member for ICLR/IJCAI/MICCAI/ECAI and a reviewer for several international journals, such as NN, INS, ESWA, KBS, SWECO and IEEE JBHI.
\end{IEEEbiography}
\vspace{-1cm}

\begin{IEEEbiography}[{\includegraphics[width=1in,height=1.25in]{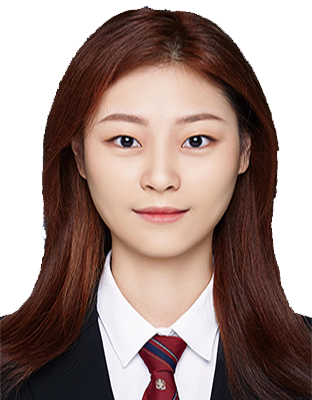}}]
{Sihan Chen}
	received the M.S. degree in electronic information from Nanjing University of Information Science and Technology, Nanjing, China, in 2024. She is currently pursuing the Ph.D. degree with the School of Computer Science and Engineering, Nanjing University of Science and Technology, Nanjing, China. Her research interests include deep learning and computer vision.
\end{IEEEbiography}

\begin{IEEEbiography}
[{\includegraphics[width=1in,height=1.25in,clip,keepaspectratio]{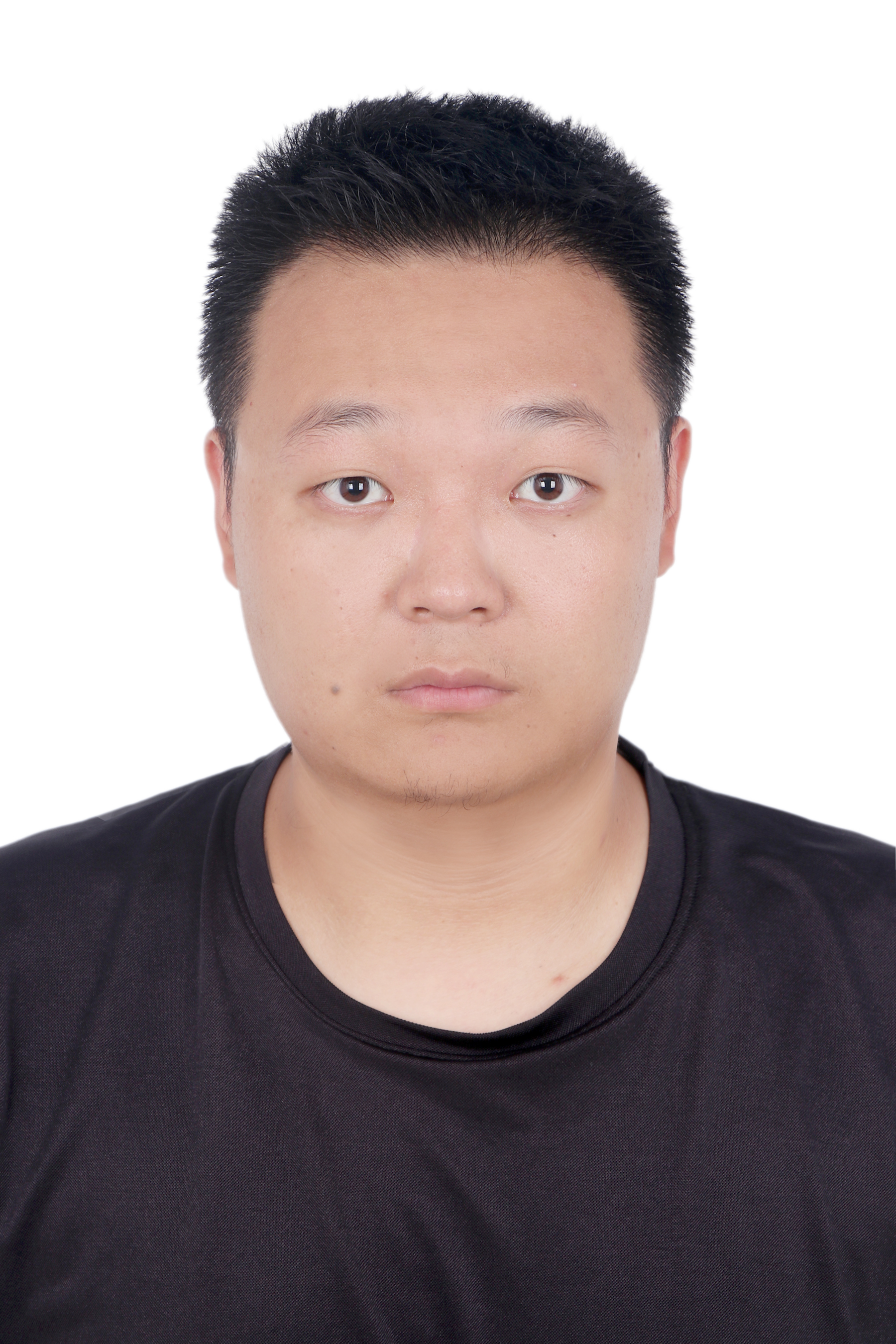}}]
{Xiantao Hu}
is currently pursuing the M.S. degree with the School of Computer Science and Engineering, Guangxi Normal University, Guilin, China.

His research interests include Tracking, Video Generation, Computer Vision.
\end{IEEEbiography}
\vspace{-1cm}
\begin{IEEEbiography}
[{\includegraphics[width=1in,height=1.25in,clip,keepaspectratio]{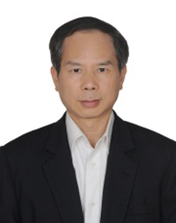}}]
{Xianxian Li}
is a Professor with the College of Computer Science and Engineering, Guangxi Normal University, Guilin, China. His research interests include data security, Internet of Things, and software theory. He has authored more than 60 refereed papers in these areas. He has served as a Program Co-Chair/Technical Program Committee Member for several IEEE conferences and workshops.
\end{IEEEbiography}

\vfill

\vspace{12pt}

\end{document}